\documentclass[aps,prb,twocolumn,groupedaddress,showpacs,floatfix]{revtex4}

\usepackage{enumerate}
\usepackage{graphics}
\usepackage{epsfig}
\usepackage{subfigure}
\usepackage{multirow}

\newcommand{\angs}{\mathrm{\AA}}

\newcommand{\be}{\begin{equation}}
\newcommand{\ee}{\end{equation}}

\newcommand{\bea}{\begin{eqnarray}}
\newcommand{\eea}{\end{eqnarray}}

\newcommand{\eqn}[1]{Eq.~(\ref{#1})}
\newcommand{\eqns}[2]{Eqns.~(\ref{#1}) and~(\ref{#2})}
\newcommand{\eqnsthree}[3]{Eqns.~(\ref{#1}),~(\ref{#2}) and~(\ref{#3})}

\newcommand{\reffig}[1]{Fig.~\ref{#1}}
\newcommand{\reffigs}[2]{Figs.~\ref{#1} and~\ref{#2}}
\newcommand{\reftab}[1]{Table~\ref{#1}}
\newcommand{\reftabs}[2]{Tables~\ref{#1} and~\ref{#2}}

\newcommand{\eform}{E_\mathrm{f}}
\newcommand{\emig}{E_\mathrm{m}}
\newcommand{\ebind}{E_\mathrm{b}}

\begin{document}

\title{Defect and solute properties in dilute Fe-Cr-Ni austenitic alloys from first principles}

\author{T. P. C. Klaver}
\email[Email: ]{klaver2@gmail.com}
\affiliation{Department of Materials Science and Engineering, Delft University of Technology, Mekelweg 2, 2628 CD Delft, The Netherlands.}
\author{D. J. Hepburn}
\email[Email: ]{dhepburn@ph.ed.ac.uk}
\author{G. J. Ackland}
\email[Email: ]{gjackland@ed.ac.uk}
\affiliation{Institute for Condensed Matter and Complex Systems, School of Physics and SUPA, The University of Edinburgh, Mayfield Road, Edinburgh, EH9 3JZ, UK.} 
\date{\today}

\begin{abstract}

We present results of an extensive set of first-principles density
functional theory calculations of point defect formation, binding and
clustering energies in austenitic Fe with dilute concentrations of Cr
and Ni solutes. A large number of possible collinear magnetic
structures were investigated as appropriate reference states for
austenite. We found that the antiferromagnetic single- and
double-layer structures with tetragonal relaxation of the unit cell
were the most suitable reference states and highlighted the inherent
instabilities in the ferromagnetic states. Test calculations for the
presence and influence of non-collinear magnetism were performed but
proved mostly negative. We calculate the vacancy formation energy to
be between 1.8 and 1.95 eV. Vacancy cluster binding was initially weak
at 0.1 eV for divacancies but rapidly increased with additional
vacancies. Clusters of up to six vacancies were studied and a highly
stable octahedral cluster and stacking fault tetrahedron were found
with total binding energies of 2.5 and 2.3 eV, respectively. The
$\langle 100\rangle$ dumbbell was found to be the most stable
self-interstitial with a formation energy of between 3.2 and 3.6 eV
and was found to form strongly bound clusters, consistent with other
fcc metals. Pair interaction models were found to be capable of
capturing the trends in the defect cluster binding energy
data. Solute-solute interactions were found to be weak in general,
with a maximal positive binding of 0.1 eV found for Ni-Ni pairs and
maximum repulsion found for Cr-Cr pairs of -0.1 eV. Solute cluster
binding was found to be consistent with a pair interaction model, with
Ni-rich clusters being the most stable. Solute-defect interactions
were consistent with Ni and Cr being modestly oversized and undersized
solutes, respectively, which is exactly opposite to the experimentally
derived size factors for Ni and Cr solutes in type 316 stainless steel
and in the pure materials. Ni was found to bind to the vacancy and to
the $\langle 100\rangle$ dumbbell in the tensile site by 0.1 eV and
was repelled from mixed and compressive sites. In contrast, Cr showed
a preferential binding to interstitials. Calculation of tracer
diffusion coefficients found that Ni diffuses significantly more
slowly than both Cr and Fe, which is consistent with the standard
mechanism used to explain radiation-induced segregation effects in
Fe-Cr-Ni austenitic alloys by vacancy-mediated diffusion. Comparison
of our results with those for bcc Fe showed strong similarity for pure
Fe and no correlation with dilute Ni and Cr.

\end{abstract}

\maketitle

\pagestyle{myheadings}
\markright{\hfill Please cite the published version of this work: Phys. Rev. B {\bf 85}, 174111 (2012).\hfill}

\section{Introduction}

Austenitic, face-centred cubic (fcc), $\gamma$-Fe based steels are key
materials in many applications. However, modelling the basic form,
$\gamma$-Fe, is challenging because it is metastable under the
zero-temperature conditions typically used in quantum mechanical
calculation.  The high temperature stabilisation of fcc over
ferromagnetic (fm) body centred cubic (bcc) $\alpha$-Fe at 1185 K is
due primarily to the onset of paramagnetism which contributes a
considerable amount of entropy in the fcc form.  Phonon and electronic
entropy also contribute, but the primacy of the magnetic effect is
underlined by the return to paramagnetic bcc $\delta$-Fe at still
higher temperature (1667 K).

There has been considerable experimental effort to stabilise
$\gamma$-Fe at lower temperatures, primarily by epitaxial growth of
thin films on Cu substrates (e.g.\ Meyerheim\cite{Meyerheim}) and by
the formation of $\gamma$-Fe precipitates by heat treatment of dilute
alloys of Fe in fcc Cu (e.g.\ Tsunoda\cite{Tsunoda} and
Hines\cite{Hines}). For a review of earlier work see Marsman and
Hafner\cite{GammaFeMarsman}. The principal motivation behind this
effort has been to study the structural and magnetic properties of
$\gamma$-Fe at temperatures low enough for stable magnetic
ordering. The study of point defects in these systems has not been
attempted.

First-principles calculations have proved to be a very reliable method
of obtaining information about radiation-induced defects, which
previously had been unreliably calculated using empirical potentials.
In non-magnetic elements such as molybdenum and vanadium it was shown
that the pseudopotential plane-wave method reproduced
experimentally-inferred self-interstitial migration energy barriers to
within 0.1 eV\cite{MoV}, giving confidence that experimentally
inaccessible quantities such as interstitial formation energies would
also be reliable. This is assisted by the discovery that the strain
fields associated with interstitials are less extensive than had been
predicted by interatomic potentials\cite{MoV,AcklandA,VDiff,EAM}, such
that calculations with supercells as small as 100 atoms can give
near-converged solutions.

Application of density functional theory (DFT) to steels is of
particular interest for radiation damage applications in which
high-energy defects such as self-interstitials are formed. Modelling
commercial steels is a more complicated task on account of their
multicomponent nature, however, there has been much progress in
ferritic Fe (e.g.\ the results of the European FP6 PERFECT
project\cite{abInitioPerfect} and references therein) and Fe-Cr
alloys\cite{FeCrOlssonA,FeCrMirzoevA,FeCrOlssonB,FeCrKlaverA,FeCrOlssonC,FeCrKlaverB}
which showed a number of unexpected outcomes. In particular, isolated
Cr atoms have a small negative heat of solution in
$\alpha$-Fe\cite{FeCrOlssonA,FeCrMirzoevA}, in apparent conflict with
the phase diagram\cite{FeCrChen} which shows a miscibility gap. This
conundrum was resolved\cite{FeCrOlssonB,FeCrKlaverA} when it was shown
that Cr atoms in bcc Fe exhibit strong nearest neighbour repulsion,
resulting from magnetic frustration, with weaker repulsion still
present up to sixth-nearest-neighbour
separation\cite{FeCrKlaverA}. Consequently, a dilute solution of Cr in
$\alpha$-Fe has a negative heat of formation only up to at most 10
at.\ \% (at 0 K), a result that is consistent with experimental results
for the heat of solution and short-range order parameter, as discussed
by Bonny {\it et al.}\cite{FeCrBonny}, but was not included in the
extrapolations present in the phase diagram. The nonlinear variation
of cohesive energy with concentration means that determining
unambiguous energies for quantities such as the binding of a
self-interstitial to a Cr solute proved impossible in concentrated
alloys\cite{FeCrKlaverB} since the calculated energy had a complex
dependence, not only on the defect, but also on concentration and the
atomic arrangement.

First-principles studies of $\gamma$-Fe with
collinear\cite{GammaFeHerper,GammaFeSpisak,GammaFeDomain,GammaFeJiang,GammaFeKong}
and non-collinear\cite{GammaFeKorling,GammaFeMarsman} magnetism have
found many distinct magnetically ordered and spin spiral (meta)stable
states lying (approximately) between 0.08 and 0.15 eV/atom above the
bcc ferromagnetic ground state, $\alpha$-Fe. However, techniques
allowing reliable first-principles calculations of the paramagnetic
state of $\gamma$-Fe are only just beginning to appear in the
literature e.g. the work of K\"{o}rmann {\it et al.} and references
therein\cite{Kormann}.  There remains some debate about whether the
paramagnetism is best represented as itinerant or involves localised
moments on the ions, however, throughout this paper we interpret our
results through the localised-moment picture. From a purely numerical
point of view there are also difficulties. The Kohn-Sham functional
applied in non-magnetic DFT has a single minimum with respect to the
wavefunctions, but collinear-magnetic DFT may have up to $2^N$ minima
for an N-atom supercell, corresponding to possible permutations of the
spin.  In practice, most of these will be unstable but one can never
be sure that the lowest energy structure has been reached. The concept
of metastability is also slippery, since the numerical algorithms used
to find the minimum electronic energy do not correspond to physical
pathways which the material can follow: The very definition of
metastability is then, to some extent, algorithm dependent.

The Born-Oppenheimer approximation is used to decouple electronic and
atomic degrees of freedom.  The status of magnetic degrees of freedom
in this approximation is debatable.  One viewpoint is that, since
magnetism is due to electrons, {\it the} Born-Oppenheimer surface is
the one corresponding to the magnetic state with the lowest energy
globally.  An alternative view, which we adopt here, is that there are
many Born-Oppenheimer surfaces, each corresponding to a given magnetic
ordering.  Many of the magnetic states in fcc Fe are sufficiently
metastable to make this a useful distinction. It should, however, be
borne in mind that DFT calculations are almost exclusively performed
using a numerical algorithm which minimises the energy of the system
with respect to the free parameters of a set of basis functions used
to represent the wavefunction. This algorithm does not correspond to
any physical trajectory which the electrons could follow (c.f. time
dependent DFT). Consequently ``local minimum'' means a minimum from
which the algorithm, in our case block Davidson\cite{Davidson}, cannot
escape in this basis set.  It does not guarantee that the spin-state
will be a local minimum in reality.

To date (and to the best of our knowledge) only a few first-principles
studies of solutes and impurities have been performed in austenitic
Fe\cite{GammaFeJiang,GammaFeKong,GavriljukA,Boukhvalov,FeCrMirzoevB,Abrikosov,GavriljukB,Nazarov}
and only the work of Nazarov\cite{Nazarov} includes defect
calculations (in order to study vacancy-hydrogen interactions). There
exists no comprehensive study of point defects and their interactions
in pure fcc Fe or in dilute solid solutions of Ni and Cr in fcc Fe. We
present here just such a study using a set of magnetically ordered
states to represent the paramagnetic state of $\gamma$-Fe. By
considering more than one magnetic state we are able to estimate the
accuracy of this assumption. This approach is certainly not ideal but
is consistent with the constraint presented by first-principles
calculations, which are performed at 0 K, where the lowest energy
states dominate. Any conclusions that we make for the paramagnetic
state of $\gamma$-Fe are clearly within the confines of this approach.

We also intend this work to provide a basis for understanding the
complex interactions present in concentrated Fe-Cr-Ni fcc alloys. We
study the relatively simple dilute case here to gain insight into the
Fe-Cr-Ni system and any extrapolations to concentrated alloys do not
include possible many body concentration dependent effects.

The layout of the paper is as follows. Section \ref{compDetails}
contains the computational details of the DFT calculations performed
in this work. In Sec. \ref{refStates} we present and discuss our
results on the bulk properties and stability of the magnetic reference
states. The energetics of point defects in pure Fe and their
interactions are discussed in Sec. \ref{pointDefects}, as is their
tendency to form small defect clusters relevant in the nucleation of
microscopic defects such as voids and dislocation loops.  The results
of dilute Ni and Cr solute calculations in defect free Fe and in
interaction with point defects are presented and discussed in
Sec. \ref{alloyCalcs}. Finally, we make our conclusions in
Sec. \ref{conclusions}.

\section{Computational Details}
\label{compDetails}

All of the following calculations have been performed using the
mainstream DFT code VASP\cite{KresseHafner,KresseFurthmuller}, a
plane-wave code that implements the projector augmented wave (PAW)
method\cite{Blochl,KresseJoubert}. Standard PAW potentials supplied
with VASP were used with exchange and correlation in the generalised
gradient approximation described by the parametrisation of Perdew and
Wang\cite{PW91} and spin interpolation of the correlation potential
provided by the improved Vosko-Wilk-Nusair
scheme\cite{vwn}. Potentials with 8, 6 and 10 valence electrons were
used for Fe, Cr and Ni, respectively.

The local magnetic moments on atoms were initialised to impose the
magnetic state ordering and were then allowed to relax. The relaxed
local magnetic moments were determined by integrating the spin density
within spheres centred on the atoms. Sphere radii of 1.302, 1.323 and
1.286 $\angs$ were used for Fe, Cr and Ni, respectively.

Calculations of the bulk properties for Fe, Cr and Ni were
performed with a sufficiently high plane-wave cutoff energy (400 eV)
and sufficiently dense $k$-point sampling of the Brillouin zone
(e.g.\ $16^3$ Monkhorst-Pack grid for a conventional bcc cell) to
ensure convergence of the energy of the system to less than 1 meV per
atom.

Elastic constants were determined numerically using stress tensor
measurements after applying small (1-2\%) strain deformations to the
equilibrium structures. These were cross-checked, where possible, with
comparable determinations using the energy. Ensuring convergence of
the elastic constants placed significantly higher demands on our
calculations than for other bulk properties. A plane-wave cutoff
energy of 600 eV was used and sufficient $k$-point sampling to ensure
that the elastic constants were converged to $\pm$ 2 GPa, e.g.\ a
$20^3$ Monkhorst-Pack grid for a conventional unit cell of the fcc
ferromagnetic low-spin (fm-LS) magnetic state. It was also necessary
to reduce the energy convergence criteria for electronic minimisation
to $10^{-8}$ eV.

Formation energies of defects and solutes were calculated in
supercells of 256 ($\pm 1$, $\pm 2$,...) atoms, with supercell
dimensions held fixed at their equilibrium values and ionic positions
free to relax. For these supercells a $2^3$ $k$-point Monkhorst-Pack
grid was used to sample the Brillouin zone. Test calculations showed
this sampling to be sufficient to converge formation energies to less
than 0.05 eV in all calculations except those involving interstitial
defects, where the uncertainty could be as high as 0.1
eV\ \cite{Supp}. Formation energy differences and binding energies
were found to be converged to less than 0.03 eV except in calculations
involving interstitial defects where the error was found to be 0.04
eV\ \cite{Supp}. These errors are sufficiently small for our
purposes. Performing calculations in a fixed supercell of volume, $V$,
results in a residual pressure, $P$, for which an Eshelby-type elastic
correction for finite size to the system energy\cite{AcklandA,HanA} of
$-P^2V/2B$, where $B$ is the bulk modulus of the bulk material, can be
applied and also serves to indicate the likely error. The size of
these corrections is negligible in much of the work presented here,
being smaller than other sources of error in our
calculations. However, in the largest of our vacancy cluster
calculation and in those containing overcoordinated defects these
finite size corrections are of relevance and are discussed in the
corresponding text.

Our data are the result of the merging of two sets of data, one
calculated in the single-layer anti-ferromagnetic (afmI) face centred
tetragonal (fct) state, the other mainly in the double layer
anti-ferromagnetic (afmD) state in either of fcc and fct structures
but also including some fct ferromagnetic high-spin (fm-HS)
calculations. The plane-wave cutoff energy was taken to be 300 eV for
the afmI calculations and 350 eV in all others. The non-convergence
error in the formation energies from either of these plane-wave cutoff
energies was found to be smaller than the $k$-point sampling
error\cite{Supp}. First-order ($N=1$) Methfessel and Paxton
smearing\cite{MethfesselPaxton} of the Fermi surface was used
throughout with smearing width, $\sigma$, set to 0.3 eV for the afmI
calculations and 0.2 eV in all others. Structural relaxations were
considered converged when the forces on all atoms were less than 0.03
eV/$\angs$ for afmI calculations and less than 0.01 eV/$\angs$ for
other magnetic states. Test calculations showed that the differences
in force convergence criteria do not lead to any significant
systematic error\cite{Supp}. The choice of $\sigma$, however, leads to
a systematic effect comparable in size to the convergence error with
respect to the number of $k$-points\cite{Supp}. This is still,
however, smaller than the uncertainty arising from choice of reference
state and is therefore sufficiently small for our purposes. Formation
energy calculations in the afmI state have been performed at the
equilibrium lattice parameters determined with those settings i.e.\ $a
= 3.4252 \angs$ and $ c = 3.648 \angs$ and not with those presented in
\reftab{elastics}. The resulting differences in formation energy are,
however, negligible.

Throughout this paper we define the formation energy,
$E_\mathrm{f}(\{n_\mathrm{X}\})$, of a configuration containing
$n_\mathrm{X}$ atoms for each element, X, relative to a set of
reference states for each element using
\begin{equation}\label{EformEq} \eform(\{n_\mathrm{X}\}) = E(\{n_\mathrm{X}\})-\sum_\mathrm{X} n_\mathrm{X}E^\mathrm{ref}_\mathrm{X}, 
\end{equation} 
where $E(\{n_\mathrm{X}\})$ is the calculated energy of the
configuration and $E^\mathrm{ref}_\mathrm{X}$ is the reference state
energy for element, X. Here we take the reference energies to be the
energies per atom for the pure materials i.e.\  Ni in its fcc fm
ground state, Cr in its bcc anti-ferromagnetic (afm) state and
Fe in the specific ground state for the magnetic ordering we are
studying.

We define the binding energy between a set of $n$ species, $\{A_i\}$,
where a species can be a defect, solute, cluster of defects and
solutes etc., as
\begin{equation}\label{EbindEq} \ebind(A_1,...,A_n)= \sum_{i=1}^n \eform(A_i) - \eform(A_1,...,A_n)
\end{equation} 
where $\eform(A_i)$ is the formation energy of a configuration
containing the single species, $A_i$, and $\eform(A_1,...,A_n)$ is the
formation energy of a configuration containing all of the species. An
energetically favoured configuration therefore has a positive binding
energy.

\section{Bulk properties and reference states}
\label{refStates}

\begin{figure}
\protect{\includegraphics[width=\columnwidth]{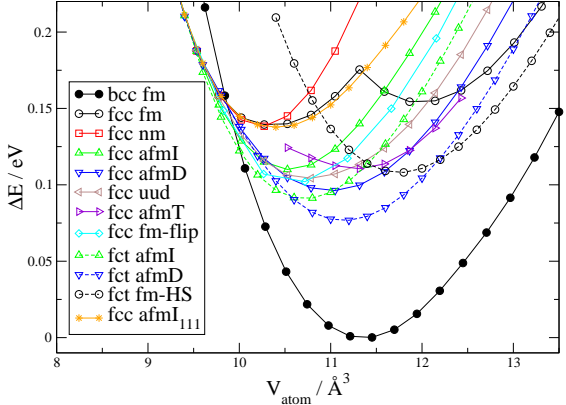}}
\caption{\label{EV_iron} Energy difference per atom, $\Delta E$,
  between distinct magnetic reference states for austenitic Fe (open
  symbols) and the bcc fm ground state (black solid circles) versus
  atomic volume, $V_\mathrm{atom}$. We include ferromagnetic (fm,black
  circles), non-magnetic (nm,red squares) and antiferromagnetic (001)
  single layer (afmI,green upward triangles), double layer (afmD,blue
  downward triangles) and triple layer (afmT,purple right facing
  triangles) orderings. Also shown is a magnetic (001) layered
  structure with a spin up, spin up, spin down ordering (uud,beige
  left facing triangles), an antiferromagnetic (111) single layer
  structure (afmI${}_{111}$, orange stars) and a magnetic ordering
  formed by taking a 4-atom fcc unit cell with three spin up atoms and
  one spin down atom (fm-flip,cyan diamonds). We distinguish between
  fcc (solid curves) and fct (dashed curves) for the same magnetic
  (001) structures.}
\end{figure}

Density functional theory has a number of energy minima corresponding
to different magnetic states with crystal structures close enough to
fcc to be plausible as reference states for austenite. We take fcc to
mean that the unit cell has $a=b=c$, in contrast with fct where the
unit cell relaxed tetragonally i.e.\  $c\ne a=b$. Our calculated energy
versus volume curves for many of these collinear magnetic structures
are shown in \reffig{EV_iron}.

The results are consistent with, and extend, previous
work\cite{GammaFeHerper,GammaFeSpisak,GammaFeDomain,GammaFeJiang,GammaFeKorling,GammaFeMarsman}. In
a similar manner to these previous studies we have concentrated on
(001) magnetic layered structures, which exhibit a common in-plane
ferromagnetism but distinct spin orderings between planes. We have,
however, considered other planar structures, such as the
afmI${}_{111}$ state, which is ferromagnetic within (111) planes but
antiferromagnetically ordered between planes, and non-planar
structures, such as the fm-flip state. While we cannot be certain we
have found the lowest energy structure within our finite dataset (as
discussed earlier), our results do show the (001) magnetic layered
structures to be generally more stable.

Overall our results show that there are many competing magnetic
structures very close in energy. Indeed, the energy difference between
the non-magnetic state and the most stable magnetic state is only
0.062 eV/atom. Test calculations in both the fcc afmI and fcc afmD
structures showed that a single flipped magnetic moment was
(meta)stable and cost 0.03 and 0.05 eV, respectively, in a 256 atom
cell, again indicating how close different magnetic structures are in
energy. Moment flips in the fcc fm-HS state proved costly at 0.5 eV
and destabilised the fcc fm-LS state with partial relaxation towards
one of the afm states. No (meta)stable moment flips were found for
any of the fct structures.

It is worth drawing the reader's attention to the fact that the afmI,
afmD and fm-HS magnetic states in cubic cells have been found to be
unstable with respect to tetragonal distortion. Both afmI and afmD
transform spontaneously when the constraint is removed but the fcc
fm-HS state is an unstable equilibrium position. Tetragonal distortion
away from the perfect fcc structure either resulted in full relaxation
to the ferromagnetic bcc ground state via the Bain path\cite{Bain},
i.e.\ by setting $c/a < 1$, or to the fct fm-HS state with $c/a >
1$. For atomic volumes below about $11.4 \angs^3$ this unstable
equilibrium position continuously becomes a stable equilibrium
position for the fcc fm-LS state. Overall the fct parameter space
exhibits many local ferromagnetic minima, the most stable being the
bcc fm ground state (for $c/a=1/\sqrt{2}$), a result most succinctly
presented in the contour plots of Spi\v{s}\'{a}k and
Hafner\cite{GammaFeSpisak}.

The elastic moduli, $\{B_{ij}\}$, of the lowest energy structures
representative of the distinct magnetic states considered here are
presented along with the lattice parameters, magnetic moments and
energy differences with respect to the lowest energy fct afmD state in
\reftab{elastics}. These elastic moduli correspond to the derivatives
of the stress tensor with respect to strains and are only equal to the
conventional elastic moduli, $\{C_{ij}\}$, which are proportional to
the second derivative of the energy with respect to strain, in the
case of a state of zero stress\cite{Wallace}. So for the fcc afmD
state, which is not in equilibrium, the $\{B_{ij}\}$ differ from the
$\{C_{ij}\}$. For all others they are identical. In the case of zero
stress the stability of the state with respect to tensile and shear
strains can be examined by calculating the eigenvalues of the elastic
constant matrix. Only if these eigenvalues are positive is the state
an energy minimum with respect to strain. For an fct structure the
eigenvalues, $\{\lambda_i\}$, are, 
\bea 
\lambda_1 & = &
C_{11} - C_{12} \nonumber \\ 
2 \lambda_2 & = & ( C_{11} + C_{12} +
C_{33} ) \nonumber \\ 
& + & \left( ( C_{11} + C_{12} - C_{33} )^2 + 8 C_{13}^2 \right)^{1/2} \nonumber \\
 2 \lambda_3 & = & ( C_{11} + C_{12} + C_{33} ) \nonumber \\ 
& - & \left( ( C_{11} + C_{12} - C_{33} )^2 + 8 C_{13}^2 \right)^{1/2} \nonumber, 
\eea 
as well as $\lambda_4 = C_{44}$, which is doubly degenerate and
$\lambda_6 = C_{66}$. The first three are given in \reftab{elastics}.

\begin{table}[htbp]
\begin{ruledtabular}
\begin{tabular}{lcccc}
Structure & fcc afmD & fct afmD & fct afmI & fct fm-HS \\
\hline
$a$/$\angs$ & 3.527 & 3.447 & 3.423 & 3.418 \\
$c$/$\angs$ & 3.527 & 3.750 & 3.658 & 4.017 \\
$c/a$ & 1.000 & 1.088 & 1.069 & 1.175 \\
$\Delta E$/eV & 0.020 & 0.000 & 0.014 & 0.031 \\
$\mu$/$\mu_B$ & 1.80 & 1.99 & 1.50 & 2.40 \\
$B_{11}$/GPa & 224 & 212 & 333 & 131 \\
$B_{12}$/GPa & 147 & 211 & 241 & 267 \\
$B_{13}$/GPa & 81 & 92 & 103 & 106 \\
$B_{31}$/GPa & 90 & 92 & 103 & 106 \\
$B_{33}$/GPa & 119 & 210 & 250 & 289 \\
$B_{44}$/GPa & 87 & 73 & 173 & 56 \\
$B_{66}$/GPa & 108 & 203 & 251 & 165 \\
$\lambda_1$/GPa & - & 1 & 92 & -136 \\
$\lambda_2$/GPa & - & 485 & 630 & 503 \\
$\lambda_3$/GPa & - & 148 & 194 & 184 \\
\end{tabular}
\end{ruledtabular}
\caption{\label{elastics} Lattice parameters, $a$ and $c$, energy per
atom relative to the fct afmD state, $\Delta E$, magnetic
moment, $\mu$, elastic moduli in Voight notation, $B_{ij}$, and the
first three eigenvalues of the matrix of elastic moduli, $\lambda_i$,
for four distinct magnetically ordered structures at their energy
minima. We estimate the uncertainties in the elastic moduli to be of
the order of a few percent. The non-zero stresses present in the fcc
afmD state are, in Voight notation, $\sigma_1 = \sigma_2 = 3~\mathrm{
GPa}$ and $\sigma_3 = -6~\mathrm{GPa}$. We measure the energy per atom
for the fct afmD state to be 0.077 eV higher than the bcc fm ground
state.}
\end{table}

The stability criterion for $\lambda_1$, i.e.\  $C_{11} > C_{12}$, is
not satisfied by the fct fm-HS state, showing that it is unstable with
respect to an orthorhombic distortion breaking the $a=b$ symmetry in
the lattice parameters. Despite being unstable it is still an
equilibrium structure i.e.\  a saddle point in the energy landscape,
which after a small orthorhombic symmetry-breaking perturbation
relaxes directly to the bcc fm ground state.  Note that although this
barrier-free double shear route from fcc to bcc (via fct) is not the
generally-considered Bain path, the relationship between the initial
and final states is identical. One final point worth mentioning here
is that the lattice parameter, $c=4.017\angs$ for the fct fm-HS state
is almost exactly $\sqrt{2}$ times the bcc fm lattice parameter, which
is 2.831~$\angs$ with the settings used here. In the Bain
transformation from fcc to bcc two of the lattice parameters must
increase to exactly this value. The conclusion is that if we only
allow one lattice parameter to increase, i.e.\  by constraining to an
fct cell with $c/a>1$, the system will still relax that lattice
parameter all the way towards bcc and not to an intermediate value.

The fct afmD state is stable, but is very soft with respect to a
further orthorhombic distortion i.e.\ upon applying the eigen-strain
associated with $\lambda_1$. By applying such a strain we confirmed
the existence of a wide minimum about zero strain with direct
measurements of $\lambda_1$ yielding values between 5 and 10 GPa. This
direct determination is still small but larger than the indirect
measurement.

As well as considerations of structural stability any prospective
reference state suitable for our purposes should be stable with
respect to the introduction of simple point defects and solutes. The
fcc fm-LS state, being unstable even to the introduction of a vacancy,
is ruled out, as is the fcc afmI state which was found to disorder in
some calculations. The fcc fm-HS state was unsurprisingly found to be
unstable to defects breaking the fcc crystal symmetry. Using the fct
fm-HS state, as will be shown, improved the situation, although it was
still unstable to defects generating orthorhombic distortions. All
other structures were found to be stable with respect to the
introduction of defects.

The final consideration in our choice of reference states is the
usefulness of our results as a solid foundation for understanding the
complex results in concentrated Fe-Cr-Ni austenitic alloys. The afmI
magnetic ordering has been found to be the only stable state for a
concentrated austenitic alloy with composition
Fe70Cr20Ni10\cite{KlaverConc}. In addition there is evidence for the
stability of a ferromagnetic state at other
concentrations\cite{Abrikosov,Majumdar}, indicating that the
ferromagnetic state may be stabilised by alloying.

Given the considerations discussed above, we have concentrated on four
main reference states for austenitic Fe and dilute Cr
and Ni alloys in this work. These are as follows:

\begin{enumerate}[i/] 

\item The fct afmI state. It exhibits stability both in structure and
against the introduction of defects and is of direct relevance in the
study of concentrated alloys.

\item The fct afmD state. It shows reasonable structural stability,
  stability against the introduction of point defects and is the
  structure with the lowest energy of all collinear magnetic
  structures found in this work and elsewhere.

\item The fct fm-HS state. Despite its instabilities it best
  represents the ferromagnetic state and insights gleaned from its
  study have relevance for concentrated alloy systems.  It also has
  the closest volume per atom to the paramagnetic state.

\item The fcc afmD state: It is stable against the introduction of
point defects and is the lowest energy fcc state. Comparison with the
fct afmD state allows the effect of tetragonal distortion on formation
and binding energies to be studied.

\end{enumerate}

It is worth reiterating that none of these structures can, by
themselves, represent high temperature austenite or concentrated
austenitic alloys. All the results in this paper must be accepted as
approximations and only when a feature is common against multiple
reference states are we confident that it is generalisable.

Before discussing our point defect and solute calculations in these
reference states we comment on the effect of non-collinear magnetism,
which is known to be present in fcc
Fe\cite{GammaFeKorling,GammaFeMarsman}. We calculated seven systems
while allowing non-collinear solutions. The systems were a vacancy,
$\langle 001\rangle$ dumbbell and an octahedral interstitial in pure
Fe, single Cr and Ni solutes, a mixed Cr-Ni $\langle 001\rangle$
dumbbell interstitial and a Ni solute next to a vacancy. The initial
moments were set in an afmI configuration, with most of the initial
moments aligned collinearly. The directions of moments on a sufficient
number of atoms were changed, however, to perturb the system away from
the collinear solution. All moments were free to rotate into
non-collinear directions in the calculations.

Six of these seven calculations converged to collinear solutions. Only
in the case of a Ni solute situated next to a vacancy were clearly
non-parallel directions of moments on atoms observed. Even then, the
energy of the non-collinear solution was only marginally lower than
for the collinear solution and by an amount below other sources of
error. Therefore, while non-collinearity was observed, it can be
omitted for our purposes.

\section{Point defects in pure Fe}
\label{pointDefects}

The basic quantities upon which all microstructural radiation-induced
damage and segregation effects depend are the energies associated with
point defect formation, interaction and diffusion i.e.\ the behaviour
of the primary damage in radiation-induced displacement cascades.  We
calculate point defect formation and binding energies in pure
$\gamma$-Fe based on our four different magnetic reference states for
austenite. It should be borne in mind that the particular magnetic
ordering and the presence of tetragonal distortion in these states
lowers their symmetry relative to the perfect fcc crystal
structure. As a result, some defects which would have been uniquely
defined in fcc have multiple configurations with different
energies. We refer to these differences as symmetry-breaking effects
and they should primarily be taken as an additional source of
uncertainty in our calculations when conclusions about austenite and
austenitic alloys are made. In order to aid the discussion of our
results we will refer to the planes of constant moment in the bulk afm
structures and the planes perpendicular to the direction of tetragonal
distortion in the fct fm-HS state by the term, ``magnetic
planes''. The results of our point defect calculations are shown in
\reftab{pointDefectsTab}.

\begin{table}[htbp]
\begin{ruledtabular}
\begin{tabular}{lcccc}
Defect & fcc afmD & fct afmD & fct afmI & fct fm-HS \\
\hline
\multirow{2}{*}{Vacancy} & 1.672 & 1.819 & 1.953 & 1.692 \\
 & -6.67 & -2.50 & -2.74 & 1.50 \\[6pt]
\multirow{2}{*}{Octa (1)} & \multirow{2}{*}{rlx (7)} & \multirow{2}{*}{rlx (7)} & 4.353 & 3.620 \\
 & & & -4.92 & -0.71 \\[6pt]
\multirow{2}{*}{Tetra uu (2)} & 3.581 & 3.864 & \multirow{2}{*}{N/A} & 3.039 \\
 & -6.04 & -2.82 & & -4.91 \\[6pt]
\multirow{2}{*}{Tetra ud (3)} & 3.332 & 3.663 & 4.322 & \multirow{2}{*}{N/A} \\
 & 0.00 & 0.00 & 0.06 & \\[6pt]
\multirow{2}{*}{$[110]$ crowdion (4)} & \multirow{2}{*}{rlx (3)} & \multirow{2}{*}{rlx (3)} & 4.799 & 3.305 \\
 & & & -5.51 & -3.10 \\[6pt]
\multirow{2}{*}{$[011]$ crowdion uu (5)} & 3.771 & 4.255 & \multirow{2}{*}{N/A} & (-1.791) \\
 & -3.53 & -3.73 & & -2.58 \\[6pt]
\multirow{2}{*}{$[01\bar{1}]$ crowdion ud (6)} & 3.874 & 4.168 & 4.818 & \multirow{2}{*}{N/A} \\
 & 0.00 & 0.00 & 0.00 & \\[6pt]
\multirow{2}{*}{$[100]$ dumbbell (7)} & 2.978 & 3.316 & 3.531 & (-3.181) \\
 & -7.79 & -4.66 & -2.96 & -1.40 \\[6pt]
\multirow{2}{*}{$[001]$ dumbbell (8)} & 2.790 & 3.195 & 3.615 & 2.416 \\
 & -9.39 & -0.17 & 4.49 & -6.27 \\[6pt]
\multirow{2}{*}{$[001]$ dumbbell (8b)} & \multirow{2}{*}{N/A} & \multirow{2}{*}{N/A} & \multirow{2}{*}{N/A} & (-2.529) \\
 & & & & -5.29 \\[6pt]
\multirow{2}{*}{$[110]$ dumbbell (9)} & 4.290 & 4.322 & 4.803 & 3.288 \\
 & -4.86 & -4.23 & -5.81 & -3.12 \\[6pt]
$[011]$ dumbbell (10) & rlx (7) & rlx (7) & rlx (7) & rlx (8b) \\[6pt]
\multirow{2}{*}{$[111]$ dumbbell (11)} & \multirow{2}{*}{rlx (3)} & \multirow{2}{*}{rlx (3)} & 4.559 & 1.919 \\
 & & & -2.13 & 0.70 \\
\end{tabular}
\end{ruledtabular}
\caption{\label{pointDefectsTab} Point defect formation energies,
  $\eform$, in eV (upper number) for the different (magnetic)
  structures studied in this work. Total magnetic moments relative to
  the bulk material, $\Delta M$, in $\mu_B$ (lower number) are given
  for completeness with all defects that can be considered as sited
  within a magnetic plane positioned in planes of positive
  moment. Interstitial defect sites (numbers 1 to 6) are shown
  in \reffig{interstitialDefectFig}. Dumbbells are identified by their
  axis direction in the coordinate system shown in the same figure
  e.g.\ a $[001]$ dumbbell has its axis along $z$ and perpendicular to
  the magnetic planes. Where the defect was found to be unstable the
  defect formed in the relaxation is identified. The negative
  formation energies for fct fm-HS (shown in brackets) are unphysical
  and stem from the instability of this reference state with respect
  to orthorhombic distortions. The positive formation energy for the
  $[001]$ dumbbell (for fct fm-HS) corresponds to a symmetrical
  starting position. The negative formation energy (identified as 8b)
  resulted from the relaxation of the $[011]$ dumbbell, which
  initially broke the $x-y$ symmetry then relaxed to $[001]$.  Eshelby
  corrections to the formation energies are less than 0.07 eV in all
  cases and at most 0.01 eV for formation energy differences.}
\end{table}

\begin{figure}
\subfigure[]{\label{interstitialDefectFig}\includegraphics[width=0.525\columnwidth]{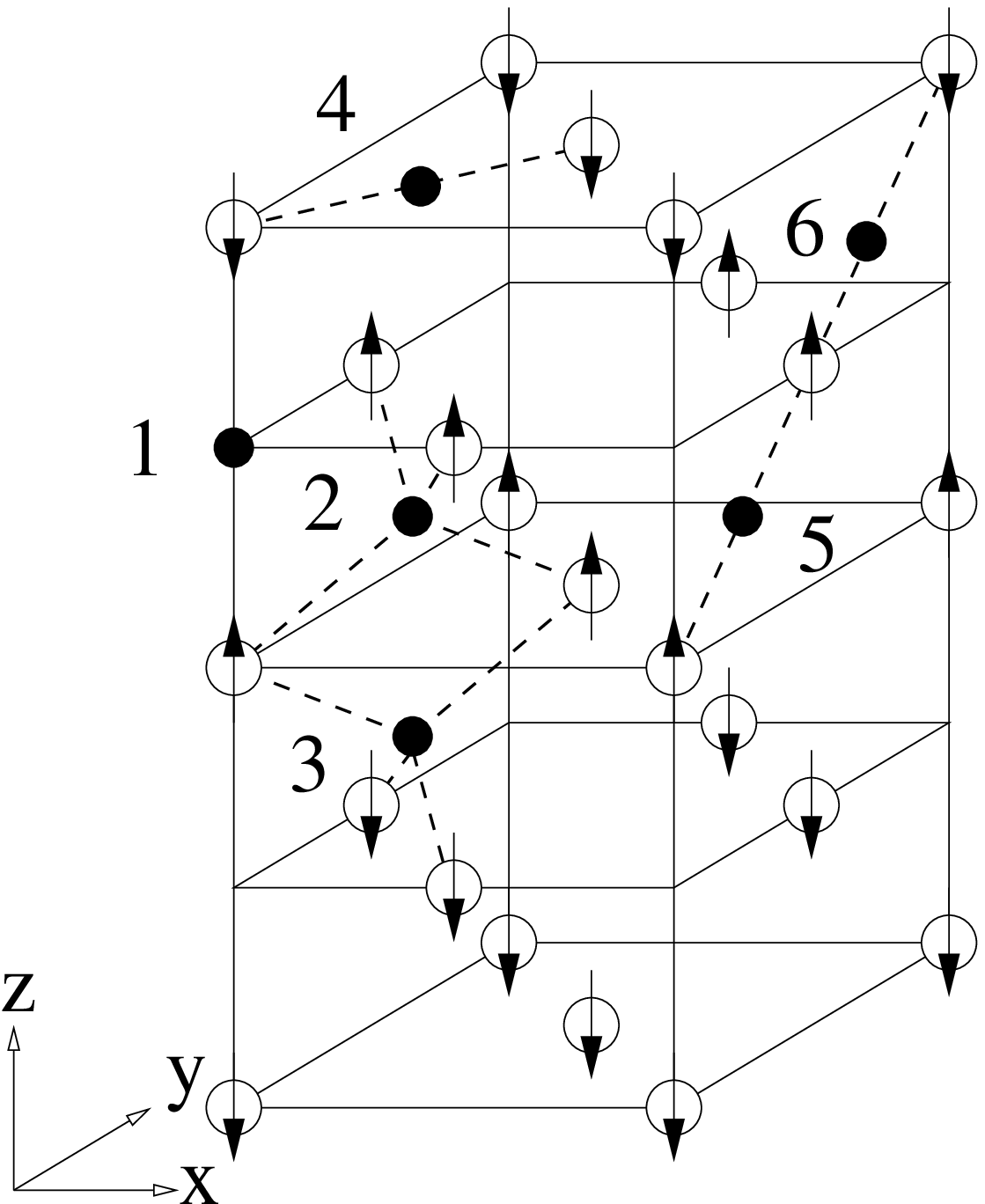}}
\subfigure[]{\label{onsiteABinteractionFig}\includegraphics[width=0.425\columnwidth]{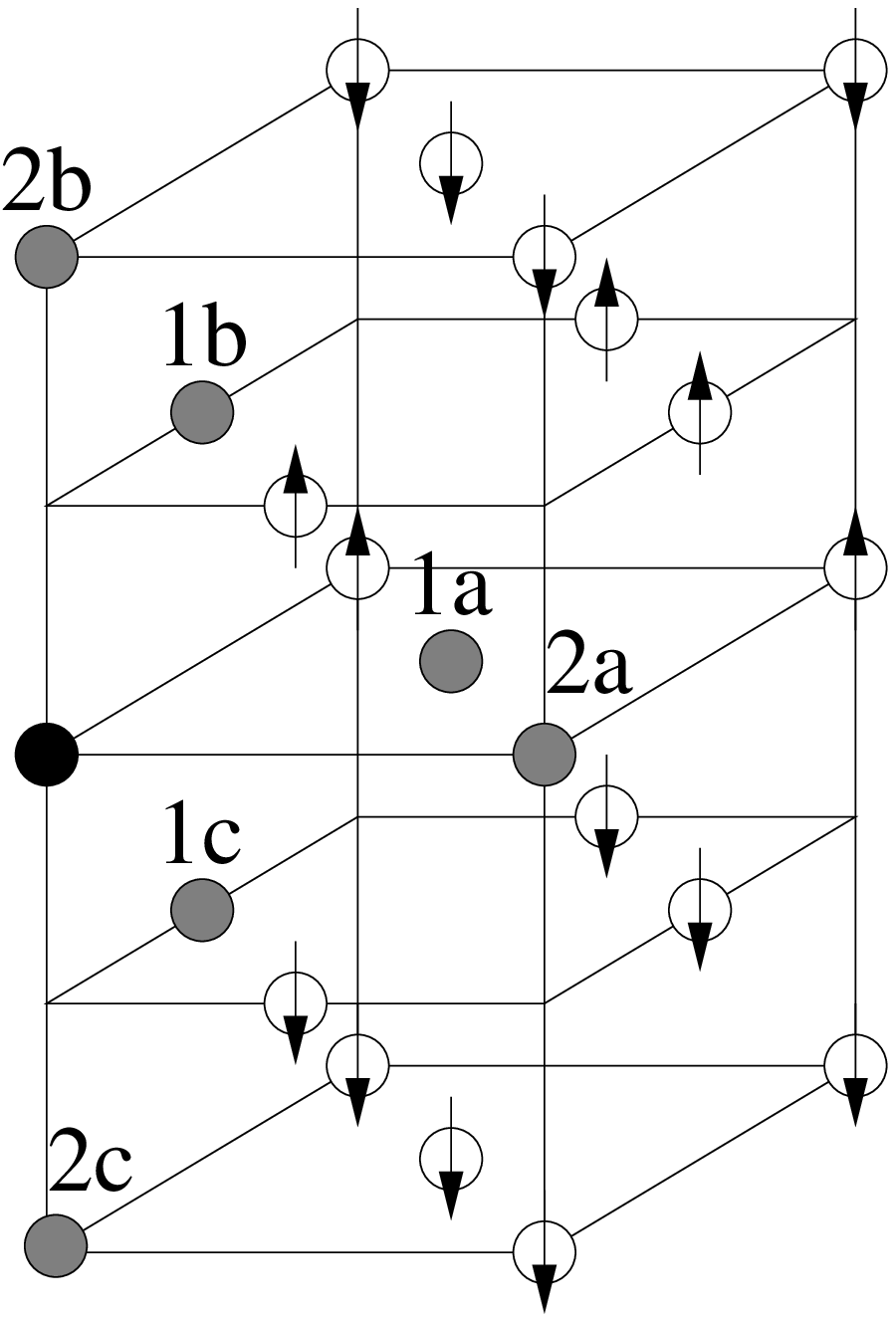}}
\caption{\label{configsFig}(a) (Self-)Interstitial defect positions
  (black circles) in the fcc/fct afmD magnetically ordered
  structure. (b) A-B interactions between on-site defects and solutes
  in the fcc/fct afmD magnetically ordered structure. Species A is
  shown in black, species B in grey and the surrounding Fe atoms in
  white. Arrows indicate local moments in both figures and the
  magnetic planes are shown to aid visualisation. The afmD state has
  been shown here to uniquely identify all of its distinct defect
  configurations. The higher symmetry afmI and fm-HS states share this
  set of configurations although many (e.g.\ tetra uu and tetra ud)
  will be symmetry equivalent.}
\end{figure}

Problems with the fct fm-HS state are immediately apparent,
i.e.\ negative interstitial formation energies. These are associated
with extensive reconstruction throughout the unit cell. All defects
exhibit significantly lower formation energies than the other
reference states. Formation energies in the fcc afmD state are
systematically lower than those in fct afmD by around 10\%. The total
configuration energies in fcc are, however, still significantly higher
than the fct state. We associate the formation energy reductions in
the fct fm-HS and fcc afmD states with the fact that these states are
not minima with respect to strain. By contrast, the fct afmI and afmD
states are minima and the influence of the defects is confined to the
first few neighbour shells. The associated energies can therefore be
regarded as attributable to the defect. We thus base our predictions
for paramagnetic austenite on these two reference states, with their
difference giving some indication of the error.

\subsection{Vacancy formation}

We find that the vacancy formation energy lies in a range between 1.82
and 1.95 eV. This is slightly higher than the typical value of one
third of the cohesive energy observed in other transition
metals\cite{Korhonen,Ehrhart,Wollenberger}. The local influence of the
vacancy on the lattice was found to be highly dependent on the
reference state. Displacements of up to 0.24~$\angs$ towards the
vacancy were found for atoms in the first nearest neighbour (1nn)
shell in the fct fm-HS state whereas no displacements exceeded
0.02~$\angs$ in the fct afmI state. An intermediate value of
0.09~$\angs$ was found in the fct afmD state. The displacements for
the fct afm states compare well with a value of 0.08~$\angs$
calculated with similar settings for fm bcc Fe. Those for the fct
fm-HS, however, appear excessive and may well be attributable to the
instability of this reference state, as discussed earlier.

Magnetic moments were typically enhanced in the 1nn shell, the effect
being strongest within a magnetic plane. Increases of 0.2~$\mu_B$ were
found for the fct afmD and fm-HS states and up to 0.35~$\mu_B$ in the
afmI state. At 2nn, moments within a magnetic plane were consistently
found to be reduced in magnitude but by less than the enhancement seen
in the 1nn shell. The overall enhancement of moments suggests that
there may be a significant magnetic contribution to the entropy of
formation for the vacancy in paramagnetic austenite.

\subsection{Interstitial formation}

All reference states give the $\langle 001\rangle$ dumbbell as the
stable self-interstitial configuration.  The fct afm states suggest a
formation energy of between 3.2 and 3.6 eV, large enough to preclude
the formation of thermal interstitials or Frenkel pairs.  Therefore,
these defects are only of importance in irradiated samples. The
symmetry-breaking effect between the [001] and [100] dumbbells is
approximately 0.1 eV i.e.\ around 3\% of the formation
energy. Experimental evidence in the fcc metals aluminium and copper
indicates the $\langle 001\rangle$ dumbbell as the most stable
self-interstitial defect\cite{Erhart}. Other
measurements\cite{Wollenberger,Schilling,Young} are consistent with
this conclusion and also suggest the same to be true of fcc Ni. In
summary all experimental evidence suggests that the $\langle
001\rangle$ dumbbell is the most stable self-interstitial in fcc
metals, with theory in strong
agreement\cite{abInitioPerfect,EAM,ATVF,deVita,Tucker} here.

Magnetic moments on the $\langle 001\rangle$ dumbbell atoms were
severely reduced or even flipped relative to the magnetic plane
containing the dumbbell in all reference states. Moment flips were
observed for the [001] dumbbell in the fm-HS state and for the [100]
dumbbell in the afmI state. In all other cases the moments were
reduced to values in the range from 0.1 to 0.5~$\mu_B$. The influence
of the $\langle 001\rangle$ dumbbell on the local lattice naturally
splits the 1nn shell into tensile sites, lying within the plane
perpendicular to the dumbbell axis, and compressive sites lying above
and below that plane. The effect on the compressive sites is by far
the most pronounced, exhibiting displacements of between 0.2 and
0.3~$\angs$ consistently for all reference states. Magnetic moments on
these atoms were consistently reduced in magnitude by at least
0.33~$\mu_B$ with a greatest reduction of 0.92~$\mu_B$ observed in the
afmI state for a [001] dumbbell. Atoms in the tensile sites at 1nn
relaxed towards the dumbbell centre by between 0.06 and 0.13~$\angs$
and exhibited enhanced moments of 0.10~$\mu_B$ in the fm-HS state and
between 0.21 and 0.44~$\mu_B$ in the afm states.

\subsection{Vacancy migration}

We present estimates of formation energies for the transition states
involved in vacancy migration and the respective barrier heights above
the energy of an isolated vacancy in \reftab{migrationTab}.

\begin{table}[htbp]
\begin{ruledtabular}
\begin{tabular}{lcccccccc}
Path & \multicolumn{2}{c}{fcc afmD} & \multicolumn{2}{c}{fct afmD} & \multicolumn{2}{c}{fct afmI} & \multicolumn{2}{c}{fct fm-HS} \\
& $\eform$ & $\emig$ & $\eform$ & $\emig$ & $\eform$ & $\emig$ & $\eform$ & $\emig$ \\
\hline
1a & 2.717 & 1.046 & 2.563 & 0.743 & 2.575 & 0.622 & 1.826 & 0.133 \\
1b & 2.384 & 0.712 & 2.867 & 1.048 & \multicolumn{2}{c}{N/A} & (-3.935 & -5.627) \\
1c & 2.940 & 1.268 & 3.401 & 1.581 & 3.677 & 1.724 & \multicolumn{2}{c}{N/A} \\
\end{tabular}
\end{ruledtabular}
\caption{\label{migrationTab} Formation energies, $\eform$, in eV for
  the transition states involved in vacancy migration and calculated
  migration barrier heights, $\emig$, in eV, calculated as the
  difference in formation energy between the transition state and the
  relaxed vacancy, as given in \reftab{pointDefectsTab}. The formation
  energies, $\eform$, are therefore equivalent to the activation
  energy for self-diffusion, $Q_0$, along each path, which is the sum
  of the vacancy formation energy and migration barrier height. The
  migration paths are labelled by the two sites involved, as given
  in \reffig{onsiteABinteractionFig}, and the transition state
  constructed with the migrating atom placed symmetrically between the
  two sites. }
\end{table}

The transition states used were those naturally suggested by symmetry
with the migrating atom placed half way between the two lattice sites
involved. Nudged elastic band (NEB) method calculations confirm this
to be the correct choice for migration within the magnetic planes of
the afmD state.

The results for the fct fm-HS state show clear signs of instability
and are included here only to illustrate this point. The broken
symmetry of the afm states means that there are several non-equivalent
barriers. For path 1c the initial and final moments on the migrating
atom are opposite in sign. The constraint of collinear magnetism means
the moment must therefore either be zero at some point along the path,
which is very likely to be the transition state in that case given the
high energy cost of suppressing the moment in Fe, or discontinuously
flip sign at some point. In our calculations a stable moment of zero
was ensured by symmetry in the transition state, resulting in
relatively large energy barriers. However, there is no reason to
constrain the moment to be continuous along the migration path and the
inclusion of discontinuous flips would very likely lower the barrier
height and the values in \reftab{migrationTab} should therefore be
taken as upper limits.

Along paths 1a and 1b the initial and final moments for the migrating
atom have the same sign and so no moment flips are required. In the
afmI state, however, the most stable moment for the migrating atom was
found to have opposite sign to the initial and final points. A stable
solution was found with the same sign of moment but this was found to
be around 0.4 eV higher in energy. Despite this complication, our
estimate of the barrier height still stands if we allow discontinuous
moment flips along the migration path. Our data show that moment flips
have an energy cost of around 0.05 eV and would therefore have no
effect on barrier height. We therefore estimate the relevant barriers
for vacancy migration in austenite to be in the range from 0.6 to 1.05
eV.  Combined with the formation energy, this gives good agreement
with the experimental activation energy for self-diffusion in
austenite which is $Q_0 = 2.945$ eV\cite{vacFediff}.

One final point worth mentioning is that the moments on the migrating
atoms in the transition state were consistently enhanced relative to
bulk to between 2.51 and 2.68~$\mu_B$, unless constrained to be
zero. These increases exceed those found for atoms 1nn to a vacancy
defect, as would be expected given the larger volume the migrating
atoms occupy.

\subsection{Point defect interactions and clustering}

The fate of an irradiated material is initially dependent on the
interactions between point defects.  Calculations of such quantities
are given in \reftabs{vacancyBindTab}{dumbbellBindTab}.  Again, the
low symmetry of the reference states necessitates calculation of many
configurations, however a clear picture emerges from this, that
vacancies bind to form divacancies with an energy of order 0.1 eV (or
in afmD up to 0.2 eV). This rather weak binding suggests that at
elevated temperature, divacancies will not be thermodynamically stable
and nucleation of voids will face a nucleation barrier. By contrast,
interstitials bind strongly into pairs, with binding energies of
around 1 eV for parallel [001] dumbbells on adjacent sites, just as
was observed in similar calculations in fcc Ni\cite{DomainNi} where a
binding of 0.97 eV was found. Such structures can form the nucleus of
dislocation loops.

These geometric conclusions can be taken as robust, given the good
agreement between the two magnetic structures.  The elastic strain
fields are still small ($\sim 0.02$ eV) for divacancies, however, for
di-interstitials the Eshelby correction lies between -0.2 and -0.3 eV.
The effect on binding energies is smaller due to a partial
cancellation of correction terms but is still significant, resulting
in between 0.1 and 0.15 eV increases. However, the stable geometry is
determined by the differences between binding energies, and these
converge much more rapidly.

\begin{table}[htbp]
\begin{ruledtabular}
\begin{tabular}{ccccccccc}
Cfg. & \multicolumn{2}{c}{fcc afmD} & \multicolumn{2}{c}{fct afmD} & \multicolumn{2}{c}{fct afmI} & \multicolumn{2}{c}{fct fm-HS} \\
& $\eform$ & $\ebind$ & $\eform$ & $\ebind$ & $\eform$ & $\ebind$ & $\eform$ & $\ebind$ \\
\hline
 1a & 3.139 & 0.205 & 3.602 & 0.037 & 3.843 & 0.063 & 3.558 & -0.173 \\
 1b & 3.288 & 0.056 & 3.512 & 0.127 & \multicolumn{2}{c}{N/A} & (-2.586 & 5.970) \\
 1c & 3.269 & 0.075 & 3.463 & 0.175 & 3.860 & 0.046 & \multicolumn{2}{c}{N/A} \\
 2a & 3.322 & 0.022 & 3.702 & -0.064 & 3.883 & 0.023 & (-2.600 & 5.984) \\
 2b & 3.423 & -0.079 & 3.657 & -0.018 & 3.996 & -0.090 & 3.348 & 0.036 \\
\end{tabular}
\end{ruledtabular}
\caption{\label{vacancyBindTab} Formation energies, $\eform$, and,
binding energies, $\ebind$, in eV for interacting vacancies at up to
second nearest neighbour separation. Configurations are labelled as in
\reffig{onsiteABinteractionFig}. }
\end{table}

\begin{table}[htbp]
\begin{ruledtabular}
\begin{tabular}{ccccc}
A-B/Config. & \multicolumn{2}{c}{fct afmD} & \multicolumn{2}{c}{fct afmI} \\
& $\eform$ & $\ebind$ & $\eform$ & $\ebind$ \\
\hline
$[001]$-$[001]$/1a & 5.541 & 0.850 & 6.423 & 0.807 \\
$[001]$-$[001]$/1b & 6.584 & -0.194 & \multicolumn{2}{c}{N/A} \\
$[001]$-$[001]$/1c & 6.339 & 0.052 & 7.301 & -0.071 \\
$[001]$-$[001]$/2a & 6.556 & -0.165 & 7.789 & -0.559 \\
$[001]$-$[001]$/2b & 6.398 & -0.007 & 7.344 & -0.114 \\
\hline
$[001]$-$[100]$/1a & 5.942 & 0.569 & 6.655 & 0.491 \\
$[001]$-$[100]$/1b & 5.629 & 0.882 & \multicolumn{2}{c}{N/A} \\
$[001]$-$[010]$/1b & \multicolumn{2}{c}{unstable} & \multicolumn{2}{c}{N/A} \\
$[001]$-$[100]$/1c & 5.867 & 0.644 & 6.570 & 0.576 \\
$[001]$-$[010]$/1c & \multicolumn{2}{c}{unstable} & \multicolumn{2}{c}{unstable} \\
$[001]$-$[100]$/2a & 6.284 & 0.227 & 6.987 & 0.159 \\
$[001]$-$[010]$/2a & 6.084 & 0.428 & 6.711 & 0.435 \\
$[001]$-$[100]$/2b & 6.101 & 0.410 & 7.017 & 0.129 \\
$[001]$-$[100]$/2c & 6.339 & 0.172 & \multicolumn{2}{c}{N/A} \\
\hline
$[100]$-$[100]$/1a & 6.417 & 0.214 & 7.085 & -0.023 \\
$[100]$-$[010]$/1a & \multicolumn{2}{c}{unstable} & \multicolumn{2}{c}{unstable} \\
$[100]$-$[100]$/1b & 5.928 & 0.703 & \multicolumn{2}{c}{N/A} \\
$[010]$-$[010]$/1b & 6.506 & 0.125 & \multicolumn{2}{c}{N/A} \\
$[100]$-$[010]$/1b & 6.120 & 0.511 & \multicolumn{2}{c}{N/A} \\
$[100]$-$[100]$/1c & 5.540 & 1.091 & 6.539 & 0.523 \\
$[010]$-$[010]$/1c & 6.451 & 0.181 & 7.161 & 0.099 \\
$[100]$-$[010]$/1c & 5.988 & 0.643 & 6.618 & 0.444 \\
$[100]$-$[100]$/2a & \multicolumn{2}{c}{unstable} &  \multicolumn{2}{c}{unstable} \\
$[010]$-$[010]$/2a & 6.760 & -0.129 & 7.667 & -0.605 \\
$[100]$-$[010]$/2a & 6.238 & 0.393 & 6.913 & 0.149 \\
$[100]$-$[100]$/2b & 6.572 & 0.059 & 7.548 & -0.486 \\
$[100]$-$[010]$/2b & 6.383 & 0.248 & 6.846 & 0.216 \\
\end{tabular}
\end{ruledtabular}
\caption{\label{dumbbellBindTab} Formation energies, $\eform$, and,
  binding energies, $\ebind$, in eV for interacting $\langle
  100\rangle$ SI dumbbells. Configurations are labelled as in
  \reffig{onsiteABinteractionFig} with the ordering of specific
  dumbbells, labelled as A and B, uniquely identifying a
  configuration.}
\end{table}

\begin{figure}
\subfigure[~Vacancy clusters]{\label{vacancyBindingFig}\includegraphics[width=\columnwidth]{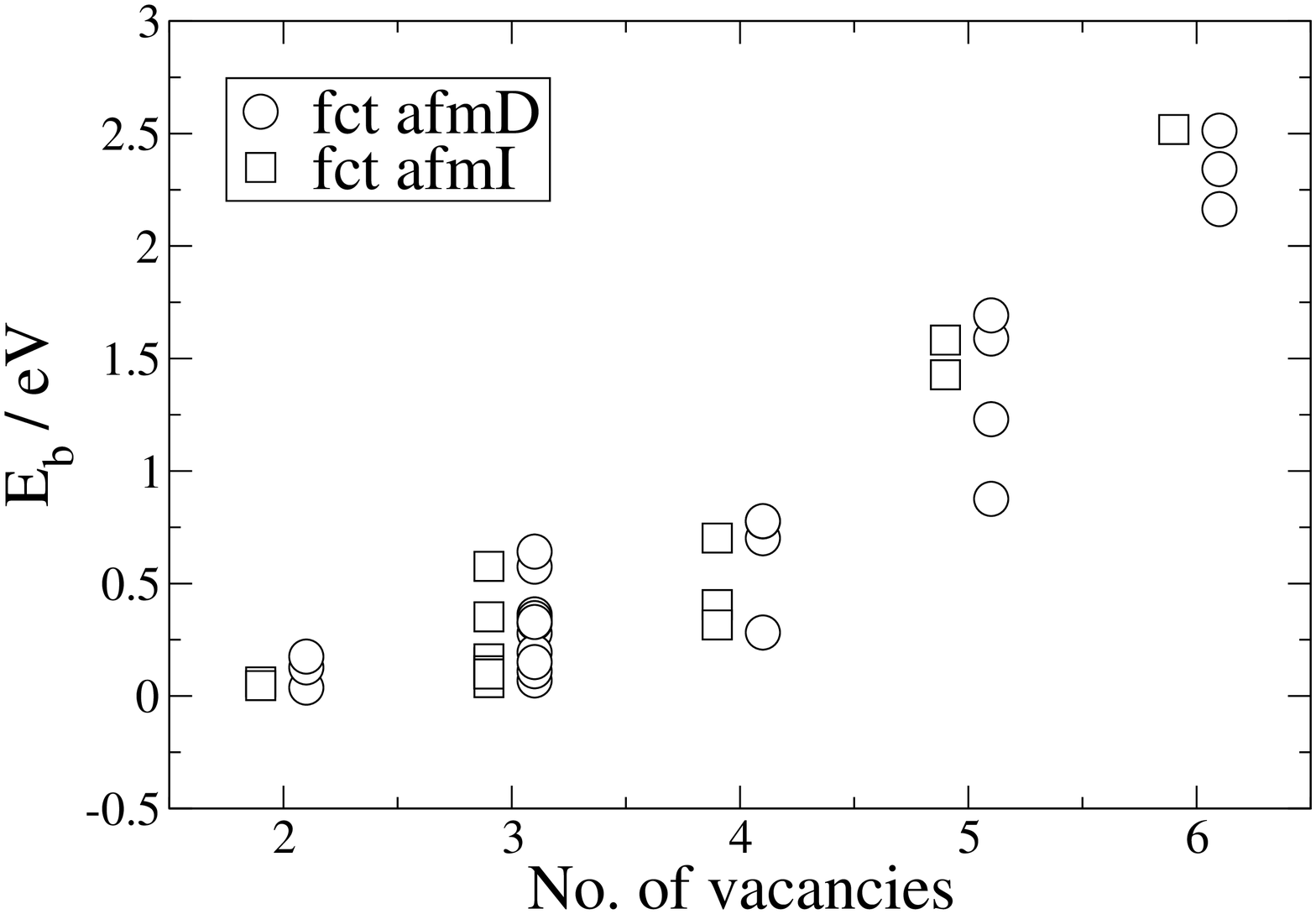}}
\subfigure[~Interstitial clusters]{\label{interstitialBindingFig}\includegraphics[width=\columnwidth]{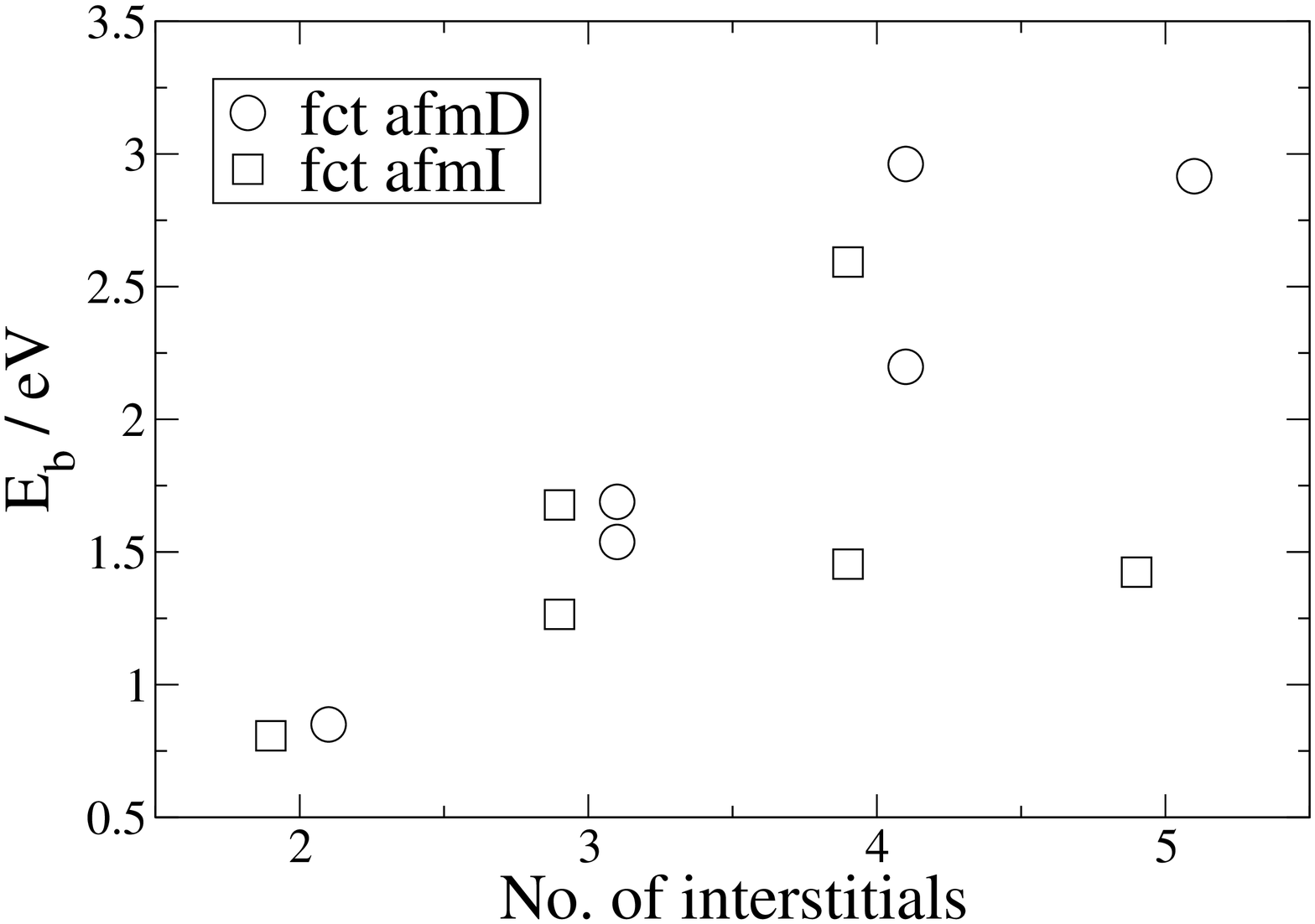}}
\caption{\label{clusterBindingFig}Total binding energies, $\ebind$, in eV
  for (a) vacancy clusters and (b) self-interstitial dumbbell
  clusters. Data have been shifted horizontally for
  clarity. Configurations of defects are considered clusters when all
  defects can be connected by chains of 1nn bonds.}
\end{figure}

To investigate void formation and dislocation loop nucleation further
we have performed a set of small defect cluster calculations. Our
choice of vacancy cluster configurations was motivated by the
observation that the strongest vacancy-vacancy binding energies are at
1nn separation (\reftab{vacancyBindTab}) and covers most of the small,
stable vacancy clusters found in other fcc metals using empirical
potential and ab initio
methods\cite{Damask,Vineyard,Shimomura,WangPot,WangAlPotDFT}. Our
results for clusters containing up to 6 vacancies are given in
\reftab{vacancyClusterTab} and presented graphically along with
results for divacancies in \reffig{vacancyBindingFig}. For
dislocation loop nucleation we consider clusters of up to 5 [001]
dumbbells (axis perpendicular to magnetic planes) lying within a
single magnetic plane, as motivated by the strong 1nn binding seen for
both afmI and afmD. The results are presented in
\reftab{001ClusterTab} and shown along with the pair binding results
in \reffig{interstitialBindingFig}.

\begin{table}[htbp]
\begin{ruledtabular}
\begin{tabular}{ccccc}
Config. & \multicolumn{2}{c}{fct afmD} & \multicolumn{2}{c}{fct afmI} \\
& $\eform$ & $\ebind$ & $\eform$ & $\ebind$ \\
\hline
(0,5,6) & 5.388 & 0.069 & 5.711 & 0.148 \\
(0,9,11) & 5.094 & 0.363 & 5.746 & 0.113 \\
(0,1,3) & 5.179 & 0.279 & \multicolumn{2}{c}{N/A} \\
(0,1,9) & 5.134 & 0.324 & 5.800 & 0.059 \\
(0,5,7) & 5.348 & 0.110 & 5.694 & 0.165 \\
(0,1,11) & 5.112 & 0.346 & 5.747 & 0.112 \\
(0,5,11) & 5.264 & 0.194 & 5.761 & 0.098 \\
(0,3,5) & 5.307 & 0.151 & \multicolumn{2}{c}{N/A} \\
(0,1,10) & 5.128 & 0.329 & 5.761 & 0.098 \\
(0,5,1) & \multicolumn{2}{c}{unstable} & \multicolumn{2}{c}{N/A} \\
(0,5,9) & \multicolumn{2}{c}{unstable} & 5.507 & 0.352 \\
(0,5,9,10) + Fe sym & 5.662 & -0.204 & 6.072 & -0.213 \\
(0,5,9,10) + Fe asym & 4.883 & 0.575 & 5.283  & 0.576 \\
(0,1,2,5) + Fe & 4.815 & 0.642 & \multicolumn{2}{c}{N/A} \\
\hline
(0,5,9,10) & 6.576 & 0.701 & 7.110 & 0.702 \\
(0,1,2,5) & 6.501 & 0.776 & \multicolumn{2}{c}{N/A} \\
(0,5,6,13) & 6.994 & 0.282 & 7.406 & 0.406 \\
(0,9,11,14) & 6.499 & 0.777 & 7.496 & 0.316 \\
\hline
(0,5,6,10,13) & 7.866 & 1.230 & 8.337 & 1.428 \\
(0,2,5,6,13) & 8.220 & 0.876 & \multicolumn{2}{c}{N/A} \\
(2,5,6,10,13) & 7.507 & 1.589 & 8.312 & 1.451 \\
(0,2,5,6,10,13)+ Fe & 7.404 & 1.692 & 8.183 & 1.582 \\
\hline
(0,2,5,6,10,13) & 8.402 & 2.513 & 9.200 & 2.518 \\
(0,1,4,5,7,16) PD & \multicolumn{2}{c}{unstable} & 10.532 & 1.184 \\
(0,1,4,5,7,16) SFT & 8.573 & 2.342 & 9.366 & 2.350 \\
(0,5,7,10,11,17) SFT & 8.751 & 2.164 & \multicolumn{2}{c}{N/A} \\
\end{tabular}
\end{ruledtabular}
\caption{\label{vacancyClusterTab}Formation, $\eform$, and total
  binding, $\ebind$, energies in eV for vacancy
  clusters. Configurations are identified by the lattice sites
  occupied by vacancies, as numbered in \reffig{clusterFig}. Each
  section of the table contains configurations with the same number of
  vacancies. Some of the configurations consist of vacancy clusters
  forming symmetrical voids with a single Fe atom placed at or near
  the centre (denoted by +Fe). In the case of the tetrahedral
  (0,5,9,10)+Fe cluster stable configurations were found with the
  central Fe atom placed symmetrically (denoted sym) and off-centre
  along an axis perpendicular to the magnetic planes (denoted
  asym). For the six vacancy clusters we distinguish between a (111)
  planar defect (PD) and a stacking fault tetrahedron (SFT) having the
  same base as the planar defect.}
\end{table}

\begin{figure}
\protect{\includegraphics[width=\columnwidth]{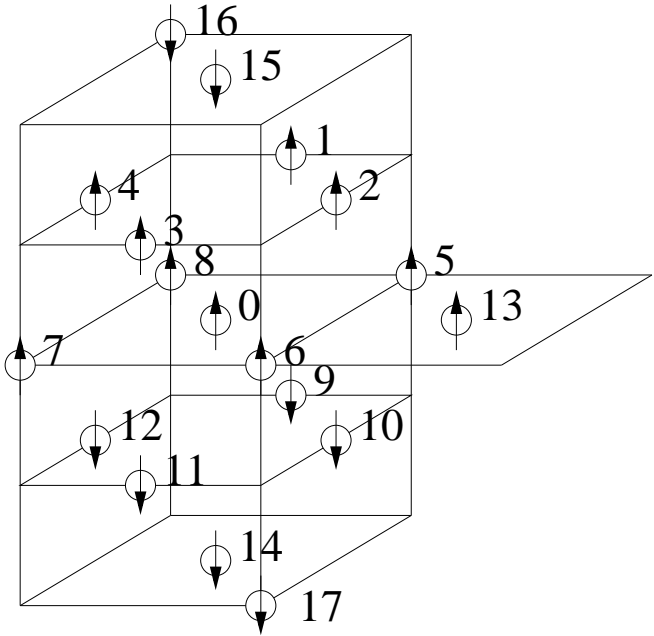}}
\caption{\label{clusterFig}Lattice site numbering used to identify
clusters of point defects and solutes. The afmD magnetic state is
shown to identify its clusters unambiguously but the numbering is
equally valid for the other magnetic states considered here (where no
ambiguity exists).}
\end{figure}

\begin{table}[htbp]
\begin{ruledtabular}
\begin{tabular}{ccccc}
Config. & \multicolumn{2}{c}{fct afmD} & \multicolumn{2}{c}{fct afmI} \\
& $\eform$ & $\ebind$ & $\eform$ & $\ebind$ \\
\hline
(0,5,6) & 8.048 & 1.538 & 9.580 & 1.265 \\
(0,5,7) & 7.897 & 1.689 & 9.168 & 1.677 \\
\hline
(0,5,6,7) & 10.583 & 2.198 & 13.006 & 1.454 \\
(0,5,6,13) & 9.820 & 2.962 & 11.867 & 2.593 \\
(5,6,7,8) & 13.055 & -0.274 & 15.628 & -1.168 \\
\hline
(0,5,6,7,8) & 13.061 & 2.916 & 16.651 & 1.424 \\
\end{tabular}
\end{ruledtabular}
\caption{\label{001ClusterTab}Formation, $\eform$, and total binding,
  $\ebind$, energies in eV for clusters of [001] SI
  dumbbells. Configurations are identified by the lattice sites
  occupied by dumbbells, as numbered in \reffig{clusterFig}.}
\end{table}

The first feature which is notable is that both vacancy and
interstitial defects have a strong tendency to bind into clusters. As
shown in \reffig{clusterBindingFig}, the total binding energy increases
steadily with the addition of defects. This is especially pronounced
for the vacancy cluster data where, despite the large spread of values
for a fixed number of vacancies, there appears to be a super-linear
increase in the total binding energy. This implies that, on average,
the binding of an additional vacancy to an already existing cluster is
increasing with cluster size. The additional binding should eventually
tend to the formation energy for a vacancy in bulk although it appears
from the data we are not at this limit yet.

The most strongly bound tri-vacancy cluster is formed by the removal
of a tetrahedron of atoms, with an additional atom replaced near the
centre. There is a strong agreement between the afmI and afmD states
here with a total binding energy of 0.58 eV and a magnetic moment on
the central atom of $3~\mu_\mathrm{B}$. Our calculations also show
that if the central Fe atom is positioned symmetrically, with zero
moment, the energy is almost 0.8 eV higher, indicating the high cost
of suppressing the Fe moment. Such a large value for the moment is
consistent with earlier observations that the local moment increases
along with the local volume occupied by an atom (unless constrained to
zero by symmetry), converging asymptotically to the moment on the free
atom. Calculations in bcc Fe are in strong agreement with this
finding\cite{bccmag}.

Significant differences between the reference states were observed for
the binding energies of the less well bound tri-vacancy clusters. A
particular case worthy of mention is configuration $(0,5,9)$
(see \reftab{vacancyClusterTab} and \reffig{clusterFig}), which is
stable in the afmI state but found to be unstable for afmD, where it
relaxes to one of the tetrahedral arrangements. Despite this, however,
the next most strongly bound tri-vacancy cluster in both cases has a
binding energy of approximately 0.35 eV, clearly indicating the
additional stability of the tetrahedral arrangement. It is well
established that stacking fault tetrahedra (SFT) are stable vacancy
configurations in fcc and this cluster could be regarded as the
smallest possible such object.

Another vacancy can be absorbed to form a tetrahedron, with an
additional binding of approximately 0.1 eV but this is significantly
lower than the binding energy per defect of the most stable
tri-vacancy so this may prove a bottleneck against forming
three-dimensional voids.

There is, however, a rapid increase in the stability of vacancy
clusters above this point. The addition of a fifth vacancy increases
the binding energy by around 0.7 eV on average. The most stable
five-vacancy configuration considered here is formed by adding a
single Fe atom to the centre of an octahedral vacancy cluster. The
local arrangement of first neighbours to this central atom is
bcc-like. There is a significant relaxation of these first neighbours
towards the central atom but the effective bcc lattice parameter is
still greater than that for bcc Fe at equilibrium and significant
tetragonal distortion remains. The moment on the central atom was
found to align with its neighbours in the afmI state i.e.\ locally
ferromagnetic with moments between 2.3 and 2.4 $\mu_B$. In afmD
similarly enhanced moments relative to the bulk were found but the
afmD magnetic order remained the stable arrangement.

The jump to six-vacancy clusters is again associated with a large
increase in total binding energy. Of particular interest is the
extremely stable octahedral arrangement of six vacancies with a total
binding energy of 2.5 eV in both the afmI and afmD states, which is
0.8 eV higher than the binding energy of any five-vacancy cluster. The
six-vacancy SFT configurations show similarly high binding
energies. These configurations are the next size of SFT up from the
elementary example discussed earlier and can be formed by removing six
atoms in a 1nn triangular arrangement from a (111) plane in fcc with a
subsequent and significant relaxation of the four atoms forming a
tetrahedron directly above this plane downward to fill the void. This
type of defect represents an alternative to the formation of
three-dimensional voids, best represented in our calculations by the
octahedral cluster. We cannot clearly distinguish between these two on
energetic grounds. One interesting aside here is that we found the
six-vacancy (111) planar defect to be metastable in the afmI state
whereas this arrangement relaxed to the SFT in afmD. This mirrors what
was found for the three-vacancy (111) planar (0,5,9) cluster. In both
of these cases, however, the SFT was found to be significantly more
stable.

The Eshelby corrections to the formation energy of most of our vacancy
cluster calculations are only a few hundredths of eV, with some
notable exceptions. Corrections for five-vacancy clusters are at most
-0.13 eV, for the octahedral cluster the correction is -0.06 eV and
for the SFT the correction is -0.2 eV. These corrections do not affect
our conclusions in any significant way.

For interstitial clusters we also find a strong tendency to bind and
align in parallel to form (001) proto-loops
(c.f. \reffig{interstitialBindingFig}), again in agreement with ab
initio results in fcc Ni\cite{DomainNi} where a tri-interstitial
binding energy of 1.71 eV was observed. The picture is complicated by
repulsion between 2nn dumbbells, which is particularly strong for the
fct afmI state. The most strongly bound planar clusters are therefore
those which maximise the ratio of 1nn to 2nn bonds. The precise
energetics and geometry of these results should be treated with care,
since our results sample a few of the very large number of possible
configurations.  Furthermore, the Eshelby correction to the binding
energy was found to be around -1 eV for the largest cluster (as
confirmed by a direct constant pressure, $P = 0~\mathrm{GPa}$,
calculation), a value that is comparable with the binding itself.
Even so, this correction can only increase the binding energy,
enhancing the driving force for clustering.

\subsection{Pair bond models for defect clusters}

We have attempted to model the binding energy of pairs of defects and
defect clusters with a linear pairwise bonding model up to
second-nearest neighbours,
\be 
\ebind^\mathrm{(m)}(n_1,n_2) = p_1 n_1 + p_2 n_2, 
\ee 
where $n_1$ and $n_2$ are the total number of 1nn and 2nn bonds,
respectively, between defects in the cluster and $p_1$ and $p_2$ are
the corresponding fit parameters. This model does not distinguish
between the distinct neighbour bonds resulting from the
symmetry-breaking effects of the magnetic states and tetragonal
distortion and fits using this model therefore provide an effective
averaging over distinct bonds. Including these symmetry-breaking
effects was found to significantly improve the agreement between model
and data but at the expense of considerably more parameters.

\begin{figure}
\vspace{-30pt}
\subfigure[~Vacancy cluster]{\label{vacancyClusterModelFig}\includegraphics[width=\columnwidth]{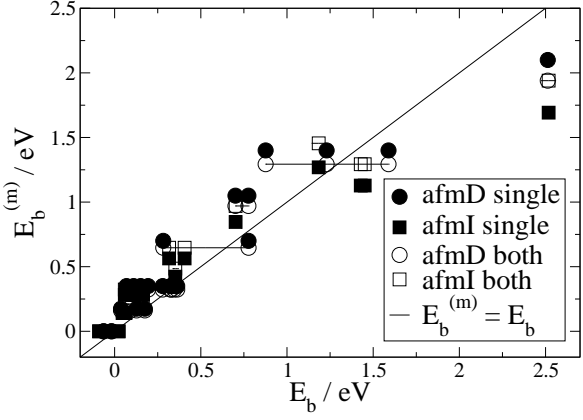}}
\subfigure[~Vacancy cluster zoomed]{\label{vacancyClusterModelFigZoom}\includegraphics[width=\columnwidth]{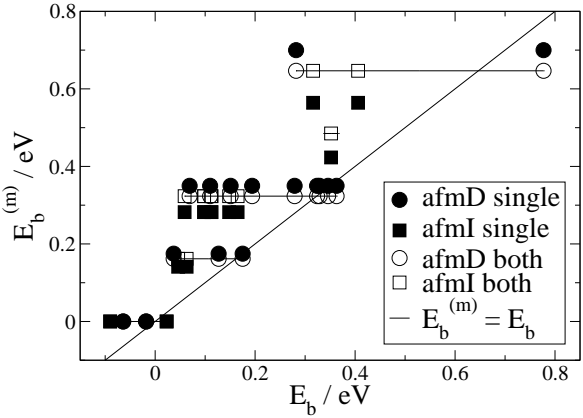}}
\subfigure[~Dumbbell cluster]{\label{dumbbellClusterModelFig}\includegraphics[width=\columnwidth]{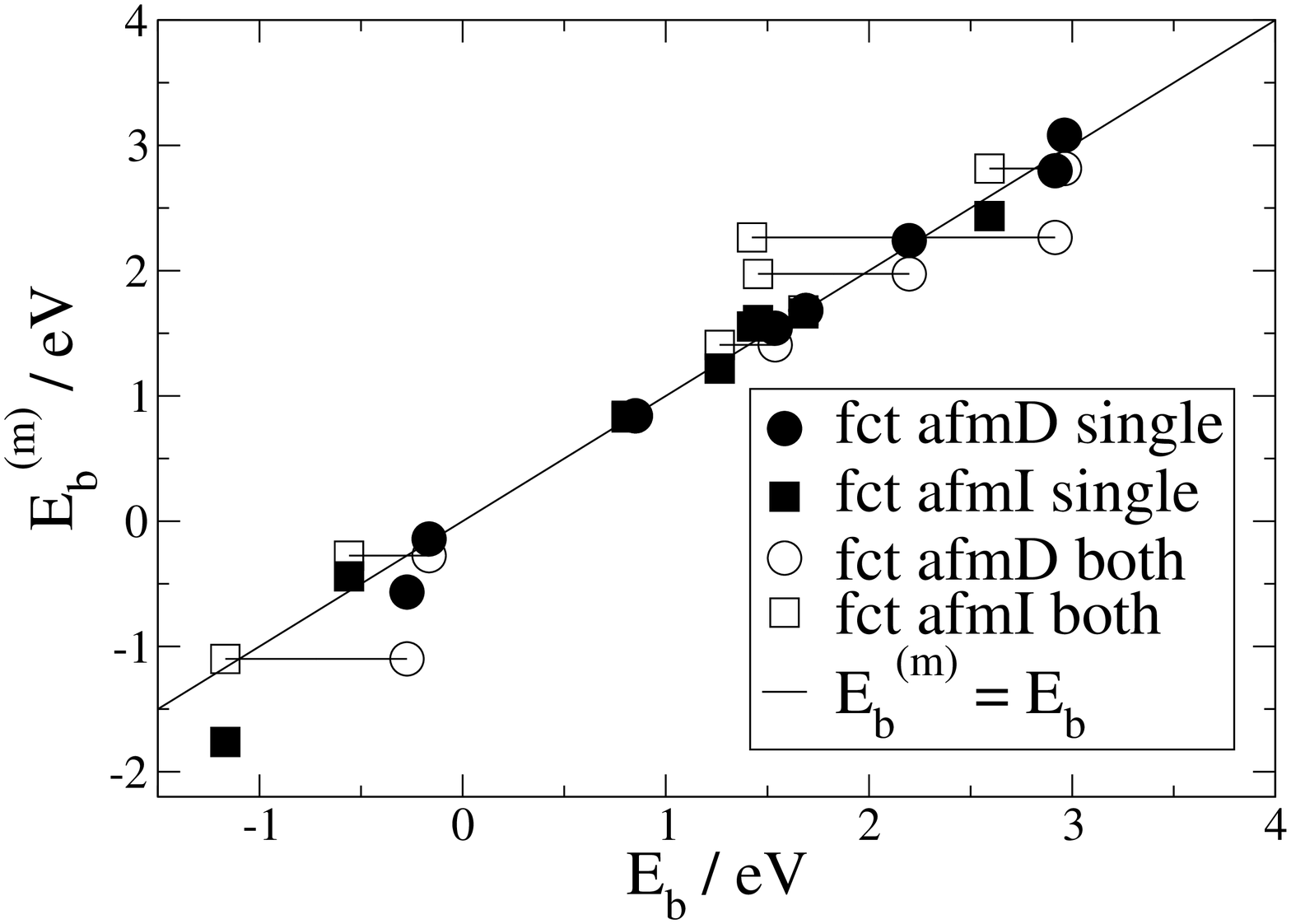}}
\caption{\label{defectClusterModelFig} Plots of binding energies from
  the pairwise bond model, $\ebind^\mathrm{(m)}$, versus ab initio fit
  data, $\ebind$, for defect clusters in the fct afmI and afmD
  states. The results of fits performed to data from a single magnetic
  state or simultaneously to both are shown, as labelled. The
  simultaneous fit results have been separated into those for afmI and
  afmD states in order to allow the effect of a combined fit to be
  directly compared with the single fit results. The
  $\ebind^\mathrm{(m)} = \ebind$ line is included to indicate a
  perfect fit. Horizontal lines indicate representative error bars for
  the averaging over both magnetic states in the combined fits.}
\end{figure}

The nature of the model also means that it is only directly applicable
to clusters where the individual defects can be assigned to single
lattice sites. This is not the case for the tetrahedral and octahedral
voids with a single Fe atom at the centre and for the six-vacancy SFT
and these configurations have not been included in the modelling. Fits
were performed both to the afmI and afmD data sets individually and to
their combined data, averaging over the two sets in the process. For
vacancy clusters the inclusion of 2nn bonds in the model made little
difference to the results and have therefore been omitted (effectively
setting $p_2=0$). The resulting fit parameters are given in
\reftab{defectClusterModelTab} and the corresponding model values are
compared with the data in \reffig{defectClusterModelFig}.

\begin{table}[htbp]
\begin{ruledtabular}
\begin{tabular}{cccc}
Parameter & afmD & afmI & both \\
\hline
\multicolumn{4}{c}{Vacancy cluster results} \\
\hline
$p_1$/eV & 0.175 & 0.141 & 0.162 \\
\hline
\multicolumn{4}{c}{Self-interstitial dumbbell cluster results} \\
\hline
$p_1$/eV & 0.841 & 0.829 & 0.841 \\
$p_2$/eV & -0.142 & -0.441 & -0.275 \\
\end{tabular}
\end{ruledtabular}
\caption{\label{defectClusterModelTab} Fit parameters, $p_i$, for the
pair bond model from fits to defect cluster binding energies for
single magnetic data sets or combined fits to both data sets as
indicated.}
\end{table}

The fit parameters for these models have the units of energy and can
be thought of as effective or averaged binding energies for pairs of
defects. Comparison with the pair binding energies shows that this is
generally true, particularly for interstitials. The afmI vacancy
cluster parameter, $p_1 = 0.141 $ eV, does, however, overestimate the
pair binding but should be thought of as an effective value which best
represents the interactions in all clusters included in the fit.

From \reffigs{vacancyClusterModelFig}{vacancyClusterModelFigZoom} it
is clear that there is a significant horizontal spread in the vacancy
cluster ab initio data for a fixed number of bonds and therefore model
value, just as there was for a fixed number of vacancies in
\reffig{vacancyBindingFig}. This is a direct result of
symmetry-breaking effects in the reference states, which the model
does not incorporate and effectively averages over. There is, however,
a broadly linear trend in the data that the model is able to
capture. This is even true of the combined fit to both data sets where
the model values lie generally within the error bars resulting from
averaging over distinct bonds and magnetic states. It is worth
mentioning that for some of the data points the model values for the
combined fit were not significantly different from the single fits and
these points in the plots are obscured as a result. Despite the
general agreement of the models they do tend to overestimate the
average binding of smaller clusters and underestimate the binding of
the largest (octahedral) cluster. This is certainly attributable to
the exceedingly strong binding of the octahedral configuration which
may well define a limit on the applicability of this model since still
larger clusters are likely to be even more strongly bound.

The agreement between models and data for interstitial clusters (shown
in \reffig{dumbbellClusterModelFig}) is significantly better than for
vacancies. This is partly due to our consideration of only planar
clusters but also demonstrates the existence of clear trends in the
data. The primary difference between the two magnetic states is the
significantly larger repulsion between dumbbells at 2nn separation for
the afmI state. Data for clusters where 2nn interactions are
significant therefore shows significant spread. Model values for the
combined fit offer an effective averaging over these differences and
provide our best guess for a predictive model of planar defects in
austenite.

\subsection{Comparison with bcc fm Fe}

\begin{figure}
\includegraphics[width=\columnwidth]{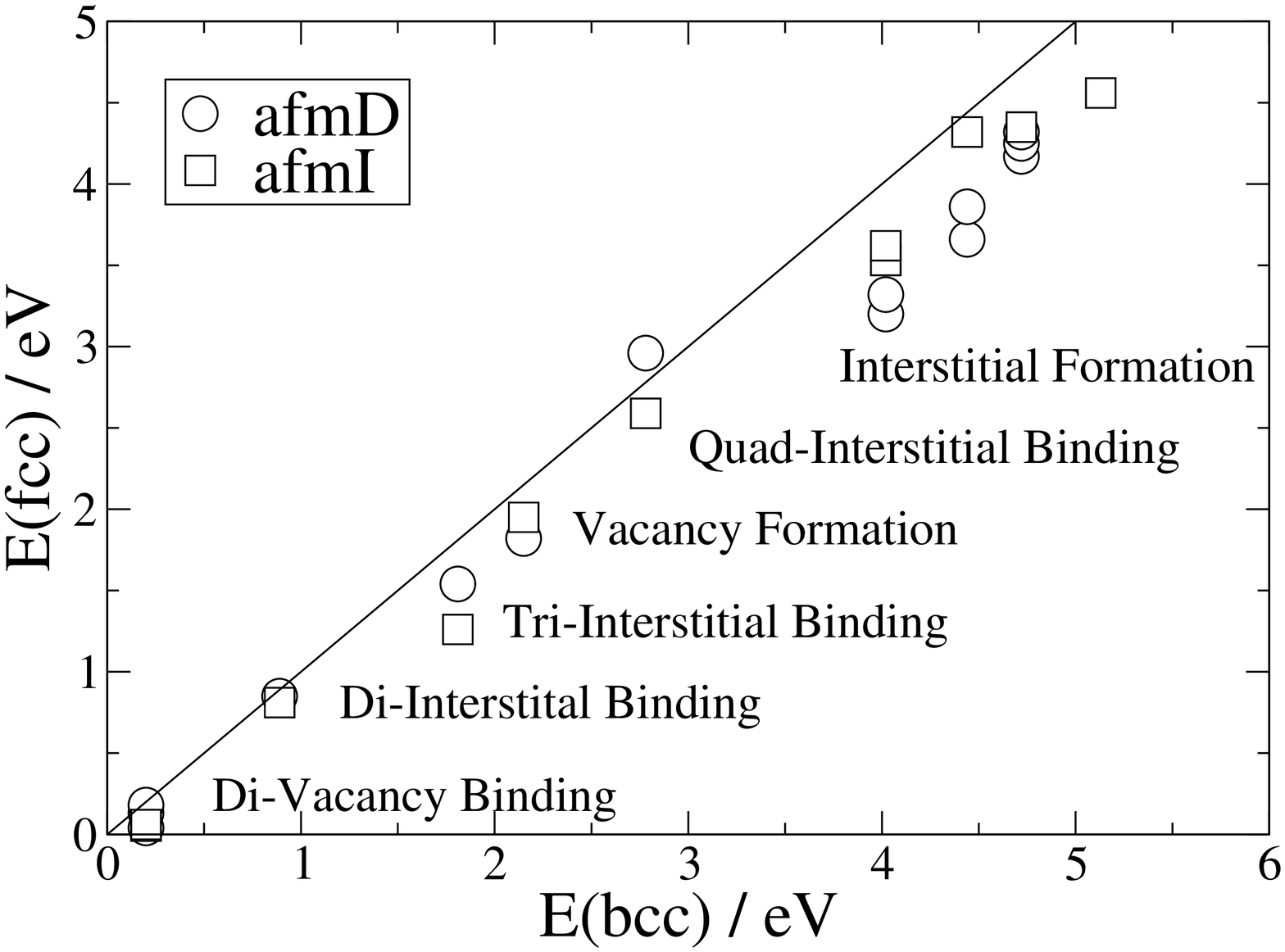}
\caption{\label{fccbccFig}Comparison of point defect formation and
  binding energies between austenitic and ferritic Fe. Interstitial
  formation energies for the two datasets have been associated with
  one another in order from lowest to highest energy. The black line
  is included to indicate an exact agreement.}
\end{figure}

As a final means to summarise our findings in pure Fe we make a
direct comparison of our data for the afm states of austenitic Fe to
the ferritic (bcc fm) ground state, as shown in \reffig{fccbccFig}.

We have included data for the formation of single vacancy and
interstitial defects\cite{FeCrOlssonC}, the divacancy binding energy
and the binding of the most stable (dumbbell) interstitial defect into
planar clusters\cite{Terentyev}. Each quantity plotted is uniquely
defined in the bcc fm state with the vertical spread in fcc values
coming from the symmetry-breaking effects in the magnetic states
considered here.

The data clearly shows a high level of similarity between the bcc and
fcc results at the energy resolution of the plot. Each quantity
occupies a clearly defined energy range for the corresponding process
e.g.\ divacancy binding from 0.0 to 0.2 eV and di-interstitial binding
from 0.8 to 0.9 eV. It also demonstrates the generally good level of
agreement between the afmI and afmD states. One noticeable difference
is the generally lower formation energies for point defects in
austenite, particularly for afmD. Despite this, the binding energies
show a generally good agreement. Overall, this raises the question of
whether such a crystal structure independence exists generally in
metals or is particular to Fe.

\section{Dilute Ni,Cr results}
\label{alloyCalcs}

Commercial austenitic stainless steels typically contain Cr and Ni as
major alloying elements. As a first step we have examined the
behaviour of these atoms in the dilute limit. Our results for single
substitutional solutes are given for the four reference states
considered here in \reftab{subSoluteTab}.

\begin{table}[htbp]
\begin{ruledtabular}
\begin{tabular}{ccccccccc}
Config. & \multicolumn{2}{c}{fcc afmD} & \multicolumn{2}{c}{fct afmD} & \multicolumn{2}{c}{fct afmI} & \multicolumn{2}{c}{fct fm-HS} \\
& $\eform$ & $\mu$ & $\eform$ & $\mu$ & $\eform$ & $\mu$ & $\eform$ & $\mu$ \\
\hline
Sub. Ni & -0.033 & -0.08 & 0.084 & 0.04 & 0.167 & -0.29 & -0.053 & 0.70 \\
Sub. Cr & 0.106 & 0.53 & 0.268 & 0.85 & 0.047 & 1.07 & 0.036 & -2.17 \\
\end{tabular}
\end{ruledtabular}
\caption{\label{subSoluteTab} Formation energies, $\eform$, in eV and
magnetic moments, $\mu$, in $\mu_B$ for substitutional Ni and Cr
solutes in austenitic Fe. The sign of the moments indicates whether there
is alignment (positive) or anti-alignment (negative) with the
moments of the atoms in the same magnetic plane.}
\end{table}

The substitutional formation energy of both Ni and Cr in austenitic
Fe is small but positive in the fct afm reference states indicating
a weak tendency for phase segregation, but only at temperatures way
below the stability limit of austenite. In the fm-HS state we find Ni
to be soluble whereas Cr is not and while the energies involved are
small this trend is opposite to that seen in similar calculations in
bcc fm Fe\cite{OlssonTMSol}. The magnetic moments on the solute
atoms exhibit the usual behaviour of Ni to be ferromagnetic to its
neighbours and Cr to be anti-ferromagnetic. In the afmD state the Ni
moment is heavily suppressed relative to its pure reference state
value ($\mu_\mathrm{Ni}^\mathrm{ref} = 0.59~\mu_B$) whereas in afmI
the moment is aligned with the majority of its 1nn Fe atoms although
still reduced in magnitude. The moment on a single Cr atom in the afm
states remains comparable to the reference state value
($\mu_\mathrm{Cr}^\mathrm{ref} = 0.89~\mu_B$) and even shows
enhancement in the afmI state. Enhanced moments are also seen for both
Ni and Cr solutes in the fm-HS state, exhibiting alignment and
anti-alignment with Fe, respectively, just as was observed in bcc fm
Fe\cite{OlssonTMSol}.

Bond lengths from single Ni and Cr solutes to their nearest-neighbour
shells differ from pure Fe by at most 0.05~$\angs$. This 1 to $2\ \%$
effect exists only for the 1nn shell and decays rapidly with
distance. A study of the magnitude and directions of bond length
changes is complicated by the symmetry-breaking effects in the
reference states. We consider instead the changes in the lattice
parameters of the unit cell surrounding a single substitutional
solute, which naturally distinguishes between effects within and
perpendicular to the magnetic planes. The build up of stresses on the
supercells was found to be consistent with the changes in the lattice
parameters.  The influence of Ni and Cr solutes was found to be very
similar within the afm reference states but showed differences in the
fm-HS state. For the fct afm states both solutes increased the lattice
parameter within a magnetic plane by 0.04~$\angs$ in afmD and
0.05~$\angs$ in afmI. The influence on the out-of-plane lattice
parameter distinguished between afmD, which showed a contraction of
0.03~$\angs$, and afmI, where an almost negligible increase was found
for Ni and a small contraction for Cr of 0.016~$\angs$. The only
significant change in the fm-HS state was a contraction of
0.04~$\angs$ in the out-of-plane lattice parameter for Ni.

The influence of Ni and Cr solutes on the magnetic moments of
surrounding Fe atoms is more pronounced. For the 1nn shell changes of
0.1~$\mu_B$ are typical but were found to be as high as 0.16~$\mu_B$
for Ni in the afmI state. No significant ($>1\ \%$) moment changes
were observed for the fm-HS state at higher orders. In the fct afmD
state moments 2nn to Ni showed perturbations of up to 0.06~$\mu_B$ and
were negligible above that. The influence of Cr was similar in
magnitude at 2nn but persisted out to the 4nn shell. In the fct afmI
state both Ni and Cr solutes exerted a more pronounced influence out
to the 4nn shell than in other states. For Ni, moments on atoms at 2nn
differed by up to 0.12~$\mu_B$ (i.e.\ $8\ \%$) from the pure Fe value
and differences of up to 0.05 $\mu_B$ were observed at 4nn. For Cr,
moment changes in the 4nn shell were the largest measured for that
separation at 0.07~$\mu_B$, exceeding those at 1nn separation, which
were up to 0.05 $\mu_B$.

Overall, we conclude that a significant contribution to the
interactions of Ni and Cr solutes with defects and other solutes in
austenitic Fe, especially at longer range, will come from their
magnetic interactions. Volume-elastic contributions will be smaller
and should be similar for Ni and Cr in the fct afmD and fct afmI
states.

\begin{table}[htbp]
\begin{ruledtabular}
\begin{tabular}{ccccccccc}
Config. & \multicolumn{2}{c}{fcc afmD} & \multicolumn{2}{c}{fct afmD} & \multicolumn{2}{c}{fct afmI} & \multicolumn{2}{c}{fct fm-HS} \\
A-B/cfg & $\eform$ & $\ebind$ & $\eform$ & $\ebind$ & $\eform$ & $\ebind$ & $\eform$ & $\ebind$ \\
\hline
Ni-Ni/1a & -0.089 & 0.024 & 0.113 & 0.056 & 0.204 & 0.110 & -0.124 & 0.018 \\
Ni-Ni/1b & -0.061 & -0.004 & 0.141 & 0.027 & \multicolumn{2}{c}{N/A} & -0.126 & 0.020 \\
Ni-Ni/1c & -0.022 & -0.043 & 0.182 & -0.014 & 0.280 & 0.034 & \multicolumn{2}{c}{N/A} \\
Ni-Ni/2a & -0.101 & 0.036 & 0.151 & 0.018 & 0.267  & 0.067 & (-5.853 & 5.748) \\
Ni-Ni/2b & -0.020 & -0.045 & 0.182 & -0.014 & 0.257 & 0.058 & -0.086 & -0.020 \\
\hline
Cr-Cr/1a & 0.257 & -0.044 & 0.562 & -0.027 & 0.173 & -0.062 & 0.313 & -0.241 \\
Cr-Cr/1b & 0.218 & -0.005 & 0.547 & -0.012 & \multicolumn{2}{c}{N/A} & (-5.742 & 5.814) \\
Cr-Cr/1c & 0.305 & -0.093 & 0.633 & -0.098 & 0.183 & -0.071 & \multicolumn{2}{c}{N/A} \\
Cr-Cr/2a & 0.170 & 0.043 & 0.512 & 0.023 & 0.086 & 0.008 & 0.102 & -0.030 \\
Cr-Cr/2b & 0.214 & -0.001 & 0.546 & -0.011 & 0.121 & -0.010 & 0.077 & -0.005 \\
\hline
Ni-Cr/1a & 0.090 & -0.016 & 0.327 & 0.025 & 0.162 & 0.051 & 0.012 & -0.028 \\
Ni-Cr/1b & 0.092 & -0.018 & 0.363 & -0.012 & \multicolumn{2}{c}{N/A} & (-5.750 & 5.734) \\
Ni-Cr/1c & 0.095 & -0.021 & 0.325 & 0.027 & 0.215 & -0.002 & \multicolumn{2}{c}{N/A} \\
Ni-Cr/2a & 0.095 & -0.021 & 0.367 & -0.016 & 0.236 & -0.022 & -0.029 & 0.012 \\
Ni-Cr/2b & 0.095 & -0.021 & 0.367 & -0.016 & 0.239 & -0.025 & 0.013 & -0.029 \\
Ni-Cr/2c & 0.098 & -0.024 & 0.359 & -0.008 & \multicolumn{2}{c}{N/A} & \multicolumn{2}{c}{N/A} \\
\end{tabular}
\end{ruledtabular}
\caption{\label{soluteSoluteTab} Formation energies, $\eform$, and,
  binding energies, $\ebind$, in eV for substitutional Ni and Cr
  solutes in Fe. Solute pair configurations are labelled according to
  the definitions in \reffig{onsiteABinteractionFig}. The afmI data
  have been produced with a $3^3$ Monkhorst-Pack $k$-point sampling,
  except for Ni-Cr 2b. Negative formation energies for the fct fm-HS
  state have once again been included (in brackets) only to illustrate
  the instability of the reference state.}
\end{table}

Interactions between pairs of Ni and Cr solutes, given in
\reftab{soluteSoluteTab}, are weak in general. The only non-negligible
attraction observed here is between pairs of Ni atoms of up to 0.1 eV
at 1nn separation and at around 0.06 eV for the afmI state at
2nn. This extended attraction may well result from the particularly
strong and long-ranging influence of Ni on neighbouring magnetic
moments in that state. Interactions between pairs of Cr atoms are
repulsive at 1nn separation and found to be particularly strong for
the fm-HS state where $\ebind = -0.24$ eV. A very similar negative
binding was found for Cr-Cr pairs in bcc fm Fe\cite{OlssonTMSol} at
1nn separation. The binding of Ni-Ni pairs was found to be almost
negligible in bcc Fe, in contrast to the modest attraction seen in afm
states here, but consistent with our findings for the fm-HS
state. Interactions between pairs of Ni and Cr are mostly negligible
save for some signs of an attraction at 1nn in the afm states. Based
on these solute pair interactions we suggest that the amount of
short-range order in alloys at typical operating temperatures for
nuclear applications will be small but with a tendency for locally
enhanced Ni-Ni and reduced Cr-Cr ordering over the random alloy.

We have performed a small set of solute cluster calculations,
containing up to five solute atoms, in order to further investigate
the trends seen for pairs of solutes and to investigate whether a
simple pair interaction model is consistent with the data. The results
of our calculations, which were performed only for the fct afm states,
are given in \reftab{soluteClusterTab}.

\begin{table}[htbp]
\begin{ruledtabular}
\begin{tabular}{ccccc}
Config. & \multicolumn{2}{c}{fct afmD} & \multicolumn{2}{c}{fct afmI} \\
& $\eform$ & $\ebind$ & $\eform$ & $\ebind$ \\
\hline
Ni:(0,5,7) & 0.141 & 0.111 & 0.272 & 0.227 \\
Ni:(0,5,6) & 0.124 & 0.128 & 0.231 & 0.269 \\
Cr:(0,5,7) & 0.795 & 0.008 & 0.266 & -0.125 \\
Cr:(0,5,6) & 0.815 & -0.012 & 0.256 & -0.115 \\
Ni:(0), Cr:(5,6) & 0.549 & 0.071 & 0.155 & 0.105 \\
Cr:(0), Ni:(5,6) & 0.361 & 0.075 & 0.202 & 0.179 \\
Ni:(5,6,14,15) & 0.336 & -0.000 & 0.430 & 0.236 \\
Cr:(5,6,14,15) & 1.085 & -0.015 & 0.251 & -0.063 \\
Ni:(0), Ni:(5,6,14,15) & 0.326 & 0.094 & 0.367 & 0.466 \\
Ni:(0), Cr:(5,6,14,15) & 1.138 & 0.016 & 0.357 & -0.003 \\
Cr:(0), Ni:(5,6,14,15) & 0.559 & 0.045 & 0.422 & 0.291 \\
Cr:(0), Cr:(5,6,14,15) & 1.386 & -0.048 & 0.422 & -0.187 \\
\end{tabular}
\end{ruledtabular}
\caption{\label{soluteClusterTab}Formation,$\eform$, and binding,
$\ebind$, energies in eV for solute clusters. Configurations are
identified by listing the lattice sites occupied by Ni and Cr solutes,
as numbered in \reffig{clusterFig}.}
\end{table}

It is immediately apparent that the most strongly bound solute
clusters contain predominantly Ni atoms. More particularly these
clusters contain a majority of Ni-Ni 1nn and 2nn pair bonds and Ni-Cr
1nn bonds indicating these results are consistent with the binding
energies of pairs of interacting solutes and that a pair interaction
model is justified. In order to quantify this claim we have performed
fits to our solute cluster and solute pair binding energies using a
pair bond model. The form of this model is similar to that used for
defect clusters but now we have to count the numbers of Ni-Ni, Ni-Cr
and Cr-Cr bonds independently, resulting in a 6 parameter model:

\bea
\ebind^\mathrm{(m)} & = & p_1^\mathrm{Ni-Ni} n_1^\mathrm{Ni-Ni} + p_2^\mathrm{Ni-Ni} n_2^\mathrm{Ni-Ni} \nonumber \\
& + & p_1^\mathrm{Cr-Cr} n_1^\mathrm{Cr-Cr} + p_2^\mathrm{Cr-Cr} n_2^\mathrm{Cr-Cr} \nonumber \\
& + & p_1^\mathrm{Ni-Cr} n_1^\mathrm{Ni-Cr} + p_2^\mathrm{Ni-Cr} n_2^\mathrm{Ni-Cr}. 
\eea

\begin{figure}
\subfigure[~Solute cluster]{\label{soluteClusterModelGraph}\includegraphics[width=\columnwidth]{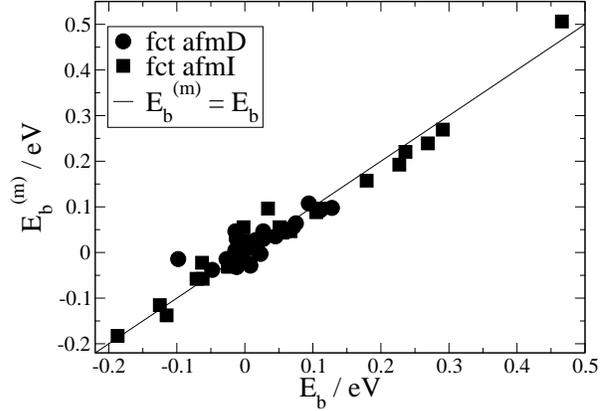}}
\subfigure[~Solute cluster zoomed]{\label{soluteClusterModelGraphZoom}\includegraphics[width=\columnwidth]{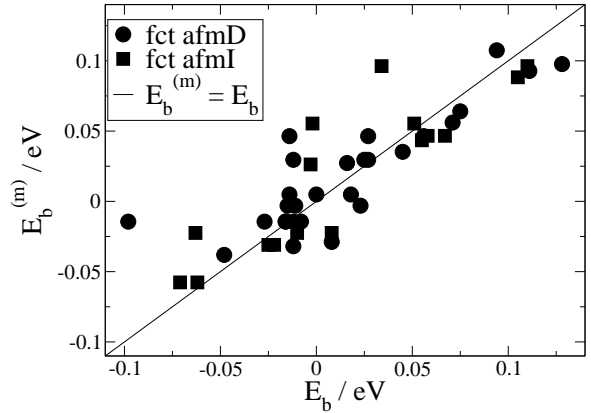}}
\caption{\label{soluteClusterModelFig} Plots of binding energies from
  the pairwise bond model, $\ebind^\mathrm{(m)}$, versus ab initio
  data, $\ebind$, for solute clusters in the fct afmI and afmD
  states. The $\ebind^\mathrm{(m)} = \ebind$ line is included to
  indicate a perfect fit.}
\end{figure}

Fits to the fct afmD dataset with this model showed good agreement
within errors. For the afmI dataset the agreement was also generally
good but there were clear outliers from the fit. Further inspection
showed that these outliers came exclusively from configurations
containing Ni atoms in the (5,6,14,15) positions i.e.\  those with Ni-Ni
3nn bonds. A calculation of the relevant Ni-Ni 3nn binding energy in
afmI gave a value of 0.055 eV. Within a pair bond model this result is
consistent with the relatively high binding energies seen in the
outliers and we have therefore included an extra term to include
contributions from Ni-Ni 3nn bonds in the afmI fit. We have not
attempted a combined fit to the afmI and afmD datasets because the
magnitude and range of the solute-solute interactions is too distinct
and the number of fit parameters differs between the sets. The results
for the 6 parameter fct afmD and 7 parameter fct afmI fits are
compared with the data in \reffig{soluteClusterModelFig} and the fit
parameters given in \reftab{soluteClusterModelTab}.

\begin{table}[htbp]
\begin{ruledtabular}
\begin{tabular}{ccc}
Parameter & afmD & afmI \\
\hline
$p_1^\mathrm{Ni-Ni}$/eV & 0.046 & 0.096 \\
$p_1^\mathrm{Cr-Cr}$/eV & -0.014 & -0.058 \\
$p_1^\mathrm{Ni-Cr}$/eV & 0.030 & 0.055 \\
$p_2^\mathrm{Ni-Ni}$/eV & 0.005 & 0.047 \\
$p_2^\mathrm{Cr-Cr}$/eV & -0.003 & -0.022 \\
$p_2^\mathrm{Ni-Cr}$/eV & -0.014 & -0.031 \\
$p_3^\mathrm{Ni-Ni}$/eV & --- & 0.043 \\
\end{tabular}
\end{ruledtabular}
\caption{\label{soluteClusterModelTab} Fit parameters,
$p_i^\mathrm{A-B}$, for the pair bond model from fits to solute
cluster binding energies for the afmI and afmD states.}
\end{table}

As was noted earlier the data presented in
\reffig{soluteClusterModelGraph} show generally good agreement between
model and data for solute clusters. The seemingly greatest
disagreement is present in the central section of the graph, as shown
in \reffig{soluteClusterModelGraphZoom}, which primarily contains
binding energies for pairs of solutes. All of the significant outlying
points, however, come from symmetry-breaking effects in the reference
states, resulting in a spread of data values corresponding to a single
model value, just as was seen for the vacancy cluster data. Our models
agree with this data in the sense that the model value for a
particular configuration lies within the spread of data, effectively
finding an average value that is most consistent with all of the
cluster data. The generally good agreement between solute pair and
cluster data strengthens our suggestion that the amount of short-range
ordering in alloys will be weak but with some tendency for enhanced
Ni-Ni ordering and the possibility of forming Ni rich clusters. Our
results do not, however, rule out the possibility of complex many-body
effects in concentrated alloys.

\subsection{Defect-Solute interaction}

We present our results for the binding of Ni and Cr solutes to a
single vacancy defect in \reftab{soluteVacTab} and for their vacancy
mediated migration in \reftab{soluteMigrationTab}.

\begin{table}[htbp]
\begin{ruledtabular}
\begin{tabular}{ccccccc}
A-B/Config. & \multicolumn{2}{c}{fcc afmD} & \multicolumn{2}{c}{fct afmD} & \multicolumn{2}{c}{fct afmI} \\
& $\eform$ & $\ebind$ & $\eform$ & $\ebind$ & $\eform$ & $\ebind$ \\
\hline
V-Ni/1a & 1.596 & 0.043 & 1.847 & 0.056 & 2.031 & 0.089 \\
V-Ni/1b & 1.626 & 0.013 & 1.876 & 0.027 & \multicolumn{2}{c}{N/A} \\
V-Ni/1c & 1.675 & -0.036 & 1.887 & 0.016 & 2.078 & 0.042 \\
V-Ni/2a & 1.614 & 0.026 & 1.906 & -0.003 & 2.109 & 0.011 \\
V-Ni/2b & 1.687 & -0.048 & 1.914 & -0.011 & 2.130 & -0.010 \\
V-Ni/2c & 1.691 & -0.051 & 1.909 & -0.005 & \multicolumn{2}{c}{N/A} \\
\hline
V-Cr/1a & 1.866 & -0.087 & 2.083 & 0.004 & 1.970 & 0.030 \\
V-Cr/1b & 1.845 & -0.066 & 2.161 & -0.075 & \multicolumn{2}{c}{N/A} \\
V-Cr/1c & 1.866 & -0.088 & 2.177 & -0.091 & 2.079 & -0.079 \\
V-Cr/2a & 1.803 & -0.025 & 2.103 & -0.017 & 2.052 & -0.052 \\
V-Cr/2b & 1.818 & -0.039 & 2.153 & -0.066 & 2.075 & -0.075 \\
V-Cr/2c & 1.814 & -0.036 & 2.091 & -0.004 & \multicolumn{2}{c}{N/A} \\
\end{tabular}
\end{ruledtabular}
\caption{\label{soluteVacTab} Formation energies, $\eform$, and, binding energies, $\ebind$, (in eV) for a vacancy defect to substitutional Ni and Cr solutes in Fe. Configuration labelling defined in \reffig{onsiteABinteractionFig}.}
\end{table}

\begin{table}[htbp]
\begin{ruledtabular}
\begin{tabular}{lcccccc}
Solute/Config. & \multicolumn{2}{c}{fcc afmD} & \multicolumn{2}{c}{fct afmD} & \multicolumn{2}{c}{fct afmI} \\
& $\eform$ & $\emig$ & $\eform$ & $\emig$ & $\eform$ & $\emig$ \\
\hline
Ni/1a & 2.901 & 1.305 & 2.738 & 0.891 & 3.014 & 0.983 \\
Ni/1b & 2.509 & 0.883 & 3.049 & 1.172 & \multicolumn{2}{c}{N/A} \\
Ni/1c & 2.608 & 0.932 & 3.066 & 1.179 & 3.441 & 1.363 \\
\hline
Cr/1a & 2.712 & 0.846 & 2.643 & 0.560 & 2.705 & 0.735 \\
Cr/1b & 2.417 & 0.572 & 2.903 & 0.742 & \multicolumn{2}{c}{N/A} \\
Cr/1c & 2.523 & 0.657 & 3.021 & 0.844 & 3.101 & 1.022 \\
\end{tabular}
\end{ruledtabular}
\caption{\label{soluteMigrationTab} Formation energies for the
  transition states, $\eform$, and calculated barrier energies,
  $\emig$, in eV for the possible vacancy migration steps involving
  solute-vacancy exchange. The transition state configurations for the
  migration are taken to be when the solute atom is midway between the
  two lattice sites involved in the migration, as labelled
  in \reffig{onsiteABinteractionFig}.}
\end{table}

We find that Ni binds to a vacancy but by no more than 0.1 eV at 1nn
and shows no sign of interaction at 2nn. In contrast the Cr-vacancy
interaction is repulsive overall, even at 2nn, with (negative) binding
energies as low as -0.091 eV. Calculations in bcc
Fe\cite{OlssonTMSol}, in contrast, show that both Cr and Ni bind to a
vacancy: Cr by 0.2 eV at 2nn separation and Ni by 0.07 eV at 1nn. The
lack of any strong tendency for vacancy binding suggests that the rate
of microstructural evolution and of creep should be relatively
unaffected, at least in dilute alloys under irradiation. There is,
however, strong experimental evidence that increasing Ni content
suppresses void formation\cite{GanWas,Allen}. The inclusion of small
quantities ($<$ 1 at.\ \%) of oversized solutes, such as Zr and Hf, in
austenitic Fe-Cr-Ni alloys was also found to significantly suppress
void formation and radiation-induced segregation (RIS) at grain
boundaries\cite{KatoVoid,KatoRIS,Allen,Ardell}. Kato\cite{KatoVoid,KatoRIS}
suggested the positive binding of vacancies to the relatively immobile
oversized solutes as a mechanism by enhancing recombination and
inhibiting vacancy diffusion, which is supported by the modelling of
Stepanov\cite{Stepanov} in the case of RIS. The vacancy-Ni binding
observed in this work is small but is likely to be cumulative and
should therefore not be overlooked as a contributory mechanism for
void suppression at higher Ni concentrations. The cumulative effect
may also be able to explain the reduction in the experimentally
determined vacancy formation energy with increasing Ni content in
FeCrNi austenitic alloys\cite{Dimitrov}.

The barrier energies for vacancy migration steps involving
solute-vacancy exchange (in \reftab{soluteMigrationTab}) show that
those for Ni are consistently higher than those for Cr by between 0.25
and 0.43 eV. In the afmD states the migration barrier heights for Fe
self-diffusion (\reftab{migrationTab}) lie consistently between those
for Ni and Cr in all but the 1c path, presumably because of the higher
cost of suppressing the Fe moment to zero, as discussed earlier. The
very same ordering of migration barrier heights was found for Cr, Fe
and self-diffusion in fcc Ni in ab initio studies by Domain and
Becquart\cite{abInitioPerfect,DomainNi} and by
Tucker\cite{Tucker,TuckerThesis} with barrier heights of 0.8, 0.95 and
1.05 eV for Cr, Fe and Ni, respectively. In the fct afmI state the low
barrier height for self-diffusion along path 1a results in a reversal
of the Fe and Cr ordering relative to this trend but Ni remains
consistently with the highest barrier.

The migration barrier heights by themselves suggest a particular
ordering for the rate of diffusion of Cr, Ni and Fe in these reference
states. However, care should be taken to incorporate correlation
effects associated with vacancy-mediated diffusion, such as those
included in the 5-frequency model of Lidiard and
LeClaire\cite{Lidiard,LeClaire}, before conclusions can be made. By
the use of suitable approximations we derive an expression for the
ratio of diffusion coefficients, $R^\mathrm{B}_\mathrm{A}$, in the
5-frequency model in Appendix \ref{appA}. This is given in
\eqns{DstarRatio}{fBapproxEq} and only depends on 4 quantities, namely
$C_\mathrm{m}$, $C_\mathrm{b}$, $H_{\mathrm{b},1}^\mathrm{B-TS}$ and
$H_{\mathrm{b},2}^\mathrm{B-TS}$, as defined in
\eqnsthree{bindEq}{CmigEq}{HeffEq}.

Our calculations allow a direct evaluation of
$H_{\mathrm{b},2}^\mathrm{B-TS}$ for each of the distinct 1nn
vacancy-solute exchange paths in the afmD and afmI states. For the
migration enthalpies, $H_{\mathrm{m},0}$ and $H_{\mathrm{m},2}$, at
zero pressure we use the migration barrier heights in
\reftabs{migrationTab}{soluteMigrationTab}, respectively. We take
$H_\mathrm{b,1nn}^\mathrm{B-V}$ at zero pressure as $\ebind$ from
\reftab{soluteVacTab}. The error associated with this use of constant
volume results for the zero-pressure case can be estimated using the
Eshelby correction term and amounts to a few meV at most. The results
for $H_{\mathrm{b},2}^\mathrm{B-TS}$ are given in
\reftab{DeltaHeffTab}.

\begin{table}[htbp]
\begin{ruledtabular}
\begin{tabular}{lccc}
Solute/Config. & fcc afmD & fct afmD & fct afmI \\
\hline
Ni/1a & -0.216 & -0.092 & -0.272 \\
Ni/1b & -0.158 & -0.097 & N/A \\
Ni/1c & 0.300 & 0.418 & 0.403 \\
\hline
Cr/1a & 0.113 & 0.187 & -0.083 \\
Cr/1b & 0.074 & 0.231 & N/A \\
Cr/1c & 0.523 & 0.646 & 0.623 \\
\end{tabular}
\end{ruledtabular}
\caption{\label{DeltaHeffTab} Results for
  $H_{\mathrm{b},2}^\mathrm{B-TS}$ in eV, as defined in \eqn{HeffEq}
  for solute vacancy exchange along the distinct 1nn paths. }
\end{table}

Results for the 1c path are clearly distinct from the others. We
attribute this to the overestimation of the self-migration barrier
along this path, resulting in an overestimation of
$H_{\mathrm{b},2}^\mathrm{B-TS}$ that is independent of the solute
species. In the afmD states, values for
$H_{\mathrm{b},2}^\mathrm{B-TS}$ along 1a and 1b are rather similar,
implying that a significant cancellation of systematic differences has
occurred in their calculation. The true value along the 1c path may
also be similar, which is consistent with their solute-independent but
reference-state-dependent overestimation. We take the arithmetic mean
of the 1a and 1b results as suitable estimates in our diffusion
coefficient modelling e.g. in the fct afmD state
$H_{\mathrm{b},2}^\mathrm{Ni-TS} = -0.094$ eV and
$H_{\mathrm{b},2}^\mathrm{Cr-TS} = 0.209$ eV. For the fct afmI state
we use the 1a results.

We expect the factors $C_\mathrm{b}$ and $C_\mathrm{m}$ to be close to
unity and weakly temperature dependent for Ni and Cr solutes in our
reference states. Ab initio evaluations of these factors for the
similar case of Cr and Fe solutes in fcc Ni from the work of Tucker
{\it et al.}\cite{Tucker} are certainly consistent with this
expectation. We set $C_\mathrm{b} = C_\mathrm{m} = 1$ in our analysis
of $R^\mathrm{B}_\mathrm{A}$ (\eqn{DstarRatio}) but can account for
any deviation from this value by noting that $R^\mathrm{B}_\mathrm{A}$
is linear in $C_\mathrm{b}$ and approximately linear (although
strictly sub-linear) in $C_\mathrm{m}$.

\begin{figure}
\subfigure[]{\label{RNiFeFig}\includegraphics[width=\columnwidth]{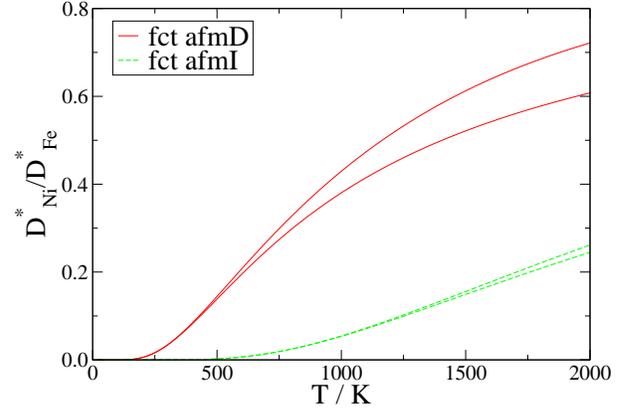}}
\subfigure[]{\label{RCrFeFig}\includegraphics[width=\columnwidth]{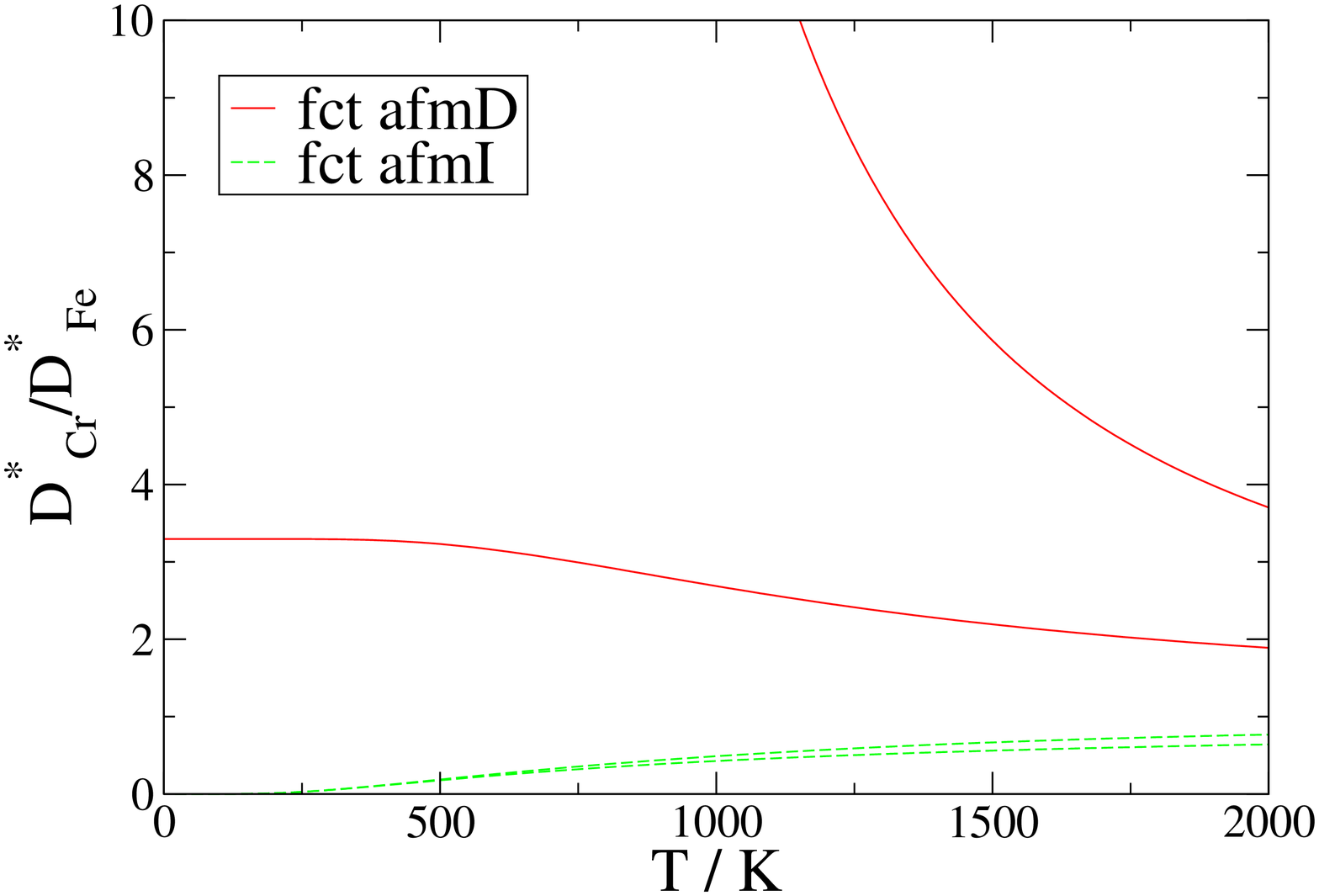}}
\subfigure[]{\label{RNiCrFig}\includegraphics[width=\columnwidth]{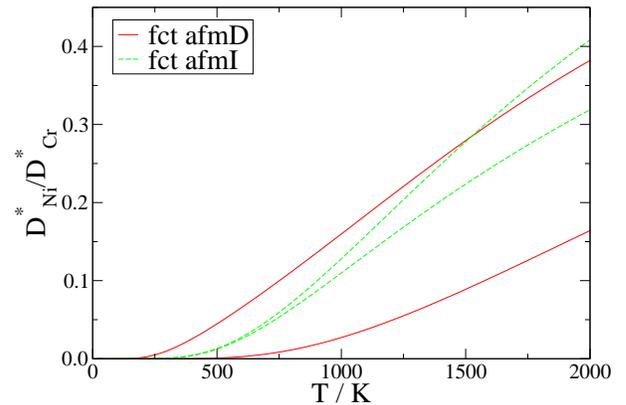}}
\caption{\label{ratioFig} Ratios of tracer diffusion coefficients
  versus temperature, $T$, for vacancy-mediated diffusion of Ni and Cr
  solutes and self-diffusion in the fct afmI and afmD Fe reference
  states. The two curves for each reference state were evaluated with
  $H_{\mathrm{b},1}^\mathrm{B-TS} = \pm 0.5$ eV and provide a
  conservative measure of the uncertainty resulting from that
  parameter. }
\end{figure}

A detailed calculation of $H_{\mathrm{b},1}^\mathrm{B-TS}$, which
would require a determination of the transition state energy using the
NEB method for a large number of symmetry nonequivalent configurations
has not been performed here. In the few cases where the transition
state is stabilised by the symmetry of the configuration, however, we
find values in the range -0.2 to 0.2 eV. We therefore conservatively
estimate that $H_{\mathrm{b},1}^\mathrm{B-TS}$ will lie somewhere
between -0.5 and 0.5 eV and since $R^\mathrm{B}_\mathrm{A}$ is a
monotonically increasing function of $H_{\mathrm{b},1}^\mathrm{B-TS}$,
its evaluation at -0.5 and 0.5 eV will give a measure of the resulting
uncertainty. We present our results for $R^\mathrm{Ni}_\mathrm{Fe}$,
$R^\mathrm{Cr}_\mathrm{Fe}$ and $R^\mathrm{Ni}_\mathrm{Cr} =
R^\mathrm{Ni}_\mathrm{Fe} / R^\mathrm{Cr}_\mathrm{Fe}$ in
\reffig{ratioFig}.

With our choice of model parameters, Ni diffusion is found to be
significantly slower than that of both Cr and Fe in both reference
states, especially at typical operating temperatures for nuclear
energy applications. The appropriate combination of $C_\mathrm{b}$ and
$C_\mathrm{m}$ factors would have to change significantly from 1
i.e. by at least a factor of 2 to alter this conclusion, which we
believe to be unlikely. The relative ordering of Cr and Fe diffusion,
however, depends on the reference state, with Cr being the fastest
diffusing species in the fct afmD state and Fe in the fct afmI
state. We cannot, therefore, make any general predictions regarding
the relative diffusivity of Cr and Fe in austenite.

The preferential association of solutes with a radiation-induced
(point) defect flux (i.e. the inverse Kirkendall effect) is posited as
the primary mechanism for RIS in Fe-Cr-Ni austenitic alloys, where Cr
depletion and Ni enhancement is observed at grain boundaries and other
defect sinks is observed\cite{Marwick,AllenRIS,WasRIS}. These
observations can be adequately explained by the preferential diffusion
of Cr over Ni by the vacancy mechanism\cite{Marwick,AllenRIS,WasRIS}
as long as the induced Cr flux is in the opposite direction to the
vacancy flux. We note, however, that interstitial mediated
diffusion\cite{Wiedersich} may also contribute to the relative rates
of diffusion, which we discuss later in this section. Our results
certainly show a preferential diffusion of Cr over Ni but an analysis
of diffusion coefficients does not determine the relative direction of
solute flow to the vacancy flux. The vacancy wind\cite{ManningWind},
$G$, which we discuss in Appendix \ref{appB}, provides a means to
investigate this question. The solute and vacancy fluxes are in
opposite directions when $G>-1$ and the same direction when $G<-1$. In
the 5-frequency model, with our approximations, the only parameter of
$G$ is $H_{\mathrm{b},1}^\mathrm{B-TS}$ (\eqn{vacancyWindEq}). When
$H_{\mathrm{b},1}^\mathrm{B-TS}\le 0$ the vacancy and solute fluxes
are opposite at all temperatures. However, if
$H_{\mathrm{b},1}^\mathrm{B-TS} > 0$ then there exists a temperature
below which the solute and vacancy flux are in the same direction (see
\reffig{vacancyWindFig}). We have not performed a detailed calculation
of $H_{\mathrm{b},1}^\mathrm{B-TS}$ in this work. However, the few
high symmetry cases we were able to calculate give a consistently
negative value for Cr and either zero or positive values for Ni. This
is not conclusive but does indicate that Cr solutes will diffuse
opposite to the vacancy flux. It also indicates that Ni is more likely
to diffuse with the vacancy flux than opposite it, which would further
enhance the Ni enhancement at defect sinks.

\begin{table}[htbp]
\begin{ruledtabular}
\begin{tabular}{lcccccc}
Defect & \multicolumn{2}{c}{fcc afmD} & \multicolumn{2}{c}{fct afmD} & \multicolumn{2}{c}{fct afmI} \\
& $\eform$ & $\ebind$ & $\eform$ & $\ebind$ & $\eform$ & $\ebind$ \\
\hline
Octa Ni (1) & \multicolumn{2}{c}{unstable} & \multicolumn{2}{c}{unstable} & 4.920 & -0.400 \\
Tetra Ni uu (2) & 4.314 & -0.765 & 4.447 & -0.499 & \multicolumn{2}{c}{N/A} \\
Tetra Ni ud (3) & 3.766 & -0.467 & 4.086 & -0.338 & 5.180 & -0.691 \\
$[110]$ Ni crow. (4) & \multicolumn{2}{c}{unstable} & \multicolumn{2}{c}{unstable} & & \\
$[011]$ Ni crow. uu (5) & 4.116 & -0.378 & 4.507 & -0.167 & \multicolumn{2}{c}{N/A} \\
$[01\bar{1}]$ Ni crow. ud (6) & 4.006 & -0.164 & 4.296 & -0.044 & 5.336 & -0.370 \\
$[100]$ FeNi dumbbell & 3.416 & -0.471 & 3.717 & -0.317 & 4.112 & -0.414 \\
$[001]$ FeNi dumbbell & 3.069 & -0.311 & 3.215 & 0.065 & 4.116 & -0.334 \\
$[001]$ NiFe dumbbell & 3.097 & -0.340 & 3.469 & -0.190 & \multicolumn{2}{c}{as FeNi} \\
$[111]$ FeNi dumbbell & 3.431 & -0.132 & 3.740 & 0.007 & 4.378 & 0.111 \\
$[111]$ NiFe dumbbell & 3.701 & -0.153 & 3.989 & -0.041 & \multicolumn{2}{c}{as FeNi} \\
\hline
Octa Cr (1) & \multicolumn{2}{c}{unstable} & \multicolumn{2}{c}{unstable} & 4.177 & 0.223 \\
Tetra Cr uu (2) & 3.550 & 0.137 & 3.785 & 0.347 & \multicolumn{2}{c}{N/A} \\
Tetra Cr ud (3) & 3.305 & 0.133 & 3.580 & 0.351 & 4.305 & 0.064 \\
$[110]$ Cr crow. (4) & \multicolumn{2}{c}{unstable} & \multicolumn{2}{c}{unstable} & & \\
$[011]$ Cr crow. uu (5) & 3.740 & 0.137 & 4.166 & 0.357 & \multicolumn{2}{c}{N/A} \\
$[01\bar{1}]$ Cr crow. ud (6) & 3.691 & 0.290 & 4.025 & 0.411 & 4.606 & 0.259 \\
$[100]$ FeCr dumbbell & 3.050 & 0.034 & 3.385 & 0.198 & 3.583 & -0.005 \\
$[001]$ FeCr dumbbell & 2.933 & -0.036 & 3.184 & 0.279 & 3.592 & 0.070 \\
$[001]$ CrFe dumbbell & 2.850 & 0.047 & 3.267 & 0.196 & \multicolumn{2}{c}{as FeCr} \\
$[111]$ FeCr dumbbell & 3.414 & \multicolumn{1}{c}{---} & 3.739 & \multicolumn{1}{c}{---} & 4.387 & 0.219 \\
$[111]$ CrFe dumbbell & \multicolumn{2}{c}{unstable} &  \multicolumn{2}{c}{unstable} & \multicolumn{2}{c}{as FeCr} \\
\hline
$[100]$ NiNi dumbbell & & & 3.897 & -0.413 & 4.620 & -0.755 \\
$[100]$ NiCr dumbbell & & & 3.643 & 0.024 & 3.939 & -0.194 \\
$[100]$ CrCr dumbbell & & & 3.826 & 0.025 & 3.930 & -0.305 \\
$[001]$ NiNi dumbbell & & & 3.438 & -0.075 & 4.160 & -0.210 \\
$[001]$ NiCr dumbbell & & & 3.344 & 0.203 & 3.943 & -0.114 \\
$[001]$ CrNi dumbbell & & & 3.162 & 0.385 & \multicolumn{2}{c}{as NiCr} \\
$[001]$ CrCr dumbbell & & & 3.536 & 0.195 & 3.931 & -0.269 \\
\end{tabular}
\end{ruledtabular}
\caption{\label{singleNiTab} Formation energies, $\eform$, and binding
  energies, $\ebind$, in eV for interstitial defects containing Ni and
  Cr solutes. Binding is defined relative to non-interacting
  self-interstitial defects of the same type and substitutional solute
  atoms. The numbering of interstitial defects is as
  in \reffig{interstitialDefectFig}. Dumbbell configurations are
  identified by their direction axis and by their composition, in
  order along that axis, with their centres considered to lie on the
  lattice site identified in \reffig{interstitialDefectFig}. All
  $\langle 110\rangle$ mixed dumbbells were found to be unstable. In
  the [111] FeNi and NiFe dumbbell calculations the Fe atom relaxed to
  a tetrahedral site and the Ni to a substitutional site at 1nn to the
  Fe. Binding energies have been calculated accordingly. Calculations
  for [110] crowdions, which were unstable for afmD, were not
  performed in the afmI state. Mixed dumbbell configurations have not
  be calculated for fcc afmD.}
\end{table}

We present our results for interstitial Ni and Cr solutes
in \reftab{singleNiTab}. It is clear from the data that Ni is strongly
repelled from interstitial sites, negative binding showing a direct
preference for self-interstitial defects and substitutional Ni. A
small but positive binding was found for the FeNi [001] dumbbell but
only for the fct afmD state. While interesting, this result is at odds
with the other reference states so cannot be taken as a general
conclusion for austenite. The generally repulsive trend is also
supported by observations that in mixed dumbbell configurations the Ni
atom is generally closer to the lattice site than Fe. For the [111]
FeNi and [111] NiFe dumbbells this asymmetry is so pronounced that the
configuration must be considered as a tetrahedral Fe interstitial with
a substitutional Ni atom at 1nn separation and the binding energies
given in the table have been calculated accordingly. Although
relatively stable, these states are always less stable than mixed
$\langle 100\rangle$ dumbbells and are important only as possible
intermediate states in the rotation, migration and disassociation of
the stable interstitials.

In contrast, Cr generally shows positive binding to interstitial
sites, an effect that is particularly prevalent in the afmD states. In
mixed dumbbell configurations the trend is for Cr atoms to be farther
from the lattice site than Fe. This was particularly pronounced for
the [111] CrFe dumbbell in the afmD state which relaxed to the
tetrahedral Cr ud configuration. Once again the mixed $\langle
100\rangle$ dumbbell configurations are the most stable.

Our calculations for doubly mixed $\langle 100\rangle$ dumbbells, show
a clear distinction between the fct afmI and afmD states. We find that
NiCr dumbbells exhibit the strongest binding in afmD and while the
same configurations are the most stable for afmI the interaction is
repulsive. One result that the two states do have in common, however,
is that in NiCr dumbbells the Ni atom is generally closest to the
lattice site and the effect is more pronounced than for mixed FeNi
dumbbells.

The emerging picture here is of a general order of preference for the
different atomic species to be found within overcoordinated defects
with Cr being the most stable, followed by Fe and finally Ni. There is
evidence of the same ordering in ab initio studies of bcc fm Fe
where Cr shows significant binding to overcoordinated
defects\cite{FeCrOlssonC,FeCrKlaverB} and there is a repulsive interaction between
Ni and the most stable dumbbell\cite{OlssonTMSol}. Ab initio
studies of dilute Cr and Fe in fcc Ni\cite{DomainNi,TuckerThesis} also
show that Cr binds strongly to the $\langle 100\rangle$ dumbbell in
mixed and doubly mixed forms but Fe shows little interaction. This
preferential binding of Cr to overcoordinated defects should result in
a positive association of Cr with the interstitial defect flux in
irradiated environments\cite{Ardell,WasRIS,Wiedersich}. This is an
interesting result as it is in the opposite sense to the flux
resulting from the vacancy mechanism of the inverse Kirkendall effect.

The same interstitial mechanism was proposed for Ni as a possible
explanation for RIS in Fe-Cr-Ni austenitic alloys, e.g.\ the work of
Watanabe\cite{Watanabe} where a binding of 0.75 eV was suggested. The
inclusion of this effect in models of RIS was, however, shown to be
inconsistent with experiment\cite{AllenRIS} for binding energies from
0.1 to 1.5 eV and our results are consistent with these findings.

\begin{table}[htbp]
\begin{ruledtabular}
\begin{tabular}{ccccccc}
Config. & \multicolumn{2}{c}{fcc afmD} & \multicolumn{2}{c}{fct afmD} & \multicolumn{2}{c}{fct afmI} \\
& $\eform$ & $\ebind$ & $\eform$ & $\ebind$ & $\eform$ & $\ebind$ \\
\hline
$[001]$-Ni/1a & 2.831 & -0.074 & 3.286 & -0.006 & 3.783 & -0.001 \\
$[001]$-Ni/1b & 2.887 & -0.129 & 3.249 & 0.030 & \multicolumn{2}{c}{N/A} \\
$[001]$-Ni/1c & 2.949 & -0.191 & 3.357 & -0.078 & 3.783 & -0.001 \\
$[001]$-Ni/2a & & & 3.313 & -0.034 & 3.861 & -0.079 \\
$[001]$-Ni/2b & & & 3.291 & -0.011 & 3.723 & 0.059 \\
$[001]$-Ni/2c & & & 3.209 & 0.070 & \multicolumn{2}{c}{N/A} \\
\hline
$[100]$-Ni/1a & & & 3.484 & -0.085 & 3.769 & -0.071 \\
$[100]$-Ni/1b & & & 3.373 & 0.027 & \multicolumn{2}{c}{N/A} \\
$[010]$-Ni/1b & & & 3.453 & -0.053 & \multicolumn{2}{c}{N/A} \\
$[100]$-Ni/1c & & & 3.379 & 0.021 & 3.710 & -0.012 \\
$[010]$-Ni/1c & & & 3.402 & -0.003 & 3.726 & -0.028 \\
$[100]$-Ni/2a & & & 3.279 & 0.121 & 3.598 & 0.100 \\
$[010]$-Ni/2a & & & 3.376 & 0.024 & 3.747 & -0.049 \\
$[100]$-Ni/2b & & & 3.404 & -0.004 & 3.692 & 0.006 \\
$[100]$-Ni/2c & & & 3.382 & 0.017 & \multicolumn{2}{c}{N/A} \\
\hline
$[001]$-Cr/1a & 2.953 & -0.056 & 3.428 & 0.035 & 3.614 & 0.048 \\
$[001]$-Cr/1b & 2.859 & 0.038 & 3.274 & 0.189 & \multicolumn{2}{c}{N/A} \\
$[001]$-Cr/1c & 2.843 & 0.053 & 3.270 & 0.193 & 3.547 & 0.115 \\
$[001]$-Cr/2a & & & 3.500 & -0.037 & 3.783 & -0.121 \\
$[001]$-Cr/2b & & & 3.394 & 0.069 & 3.670 & -0.008 \\
$[001]$-Cr/2c & & & 3.467 & -0.004 & \multicolumn{2}{c}{N/A} \\
\hline
$[100]$-Cr/1a & & & 3.407 & 0.176 & 3.527 & 0.051 \\
$[100]$-Cr/1b & & & 3.471 & 0.112 & \multicolumn{2}{c}{N/A} \\
$[010]$-Cr/1b & & & 3.492 & 0.091 & \multicolumn{2}{c}{N/A} \\
$[100]$-Cr/1c & & & 3.529 & 0.054 & 3.577 & 0.001 \\
$[010]$-Cr/1c & & & 3.461 & 0.122 & 3.575 & 0.003 \\
$[100]$-Cr/2a & & & 3.570 & 0.013 & 3.526 & 0.052 \\
$[010]$-Cr/2a & & & 3.552 & 0.031 & 3.608 & -0.030 \\
$[100]$-Cr/2b & & & 3.522 & 0.061 & 3.569 & 0.009 \\
$[100]$-Cr/2c & & & 3.598 & -0.015 & \multicolumn{2}{c}{N/A} \\
\end{tabular}
\end{ruledtabular}
\caption{\label{soluteIntTab} Formation energies, $\eform$, and,
binding energies, $\ebind$, in eV for $\langle 100\rangle$ defect to
substitutional Ni and Cr solutes in Fe. Configuration labelling is
as defined in \reffig{onsiteABinteractionFig}.}
\end{table}

The formation and binding energies of single Ni and Cr solutes to
$\langle 100\rangle$ self-interstitial dumbbells at 1nn and 2nn
separation is presented in \reftab{soluteIntTab}. In both the fct afmI
and afmD states Ni is most stable in the tensile 2nn site, collinear
with the dumbbell axis, exhibiting a binding up to approximately 0.1
eV. These are also the most stable configurations found here for a
single Ni solute interacting with an overcoordinated defect, although
in the fct afmD state the FeNi [001] dumbbell is close in
energy. Although small, this binding allows Ni to act as a weak trap
that inhibits interstitial diffusion and is therefore worth
consideration in models of microstructure evolution and RIS,
especially at higher Ni concentrations. Finally, it is worth noting
that Ni is most strongly repelled from compressive sites at 1nn to the
self-interstitial dumbbell.

For Cr there is less agreement between the two magnetic states. In fct
afmD the FeCr and CrFe [001] mixed dumbbells are the most stable,
closely followed by Cr in compressive sites at 1nn to a
self-interstitial dumbbell. The same configurations are found to be
among the most stable in the fct afmI state although others are at
least as stable with no discernible preference for tensile or
compressive sites. Despite these differences our dilute results show
that, on average, Cr exhibits positive binding to $\langle 100\rangle$
dumbbells in mixed or neighbouring sites, where like Ni it would act
as a weak trap for interstitial diffusion.

Overall, the interactions of Ni and Cr solutes with point defects in
austenitic Fe are consistent with what would intuitively be expected
of moderately oversized and undersized solutes, respectively. Ni binds
to the vacancy and is generally repelled from the mixed and
compressive sites of $\langle 100\rangle$ dumbbells but shows positive
binding in the tensile sites.  Cr generally exhibits the opposite
tendencies. Such a conclusion should, however, be contrasted with the
lack of consistent behaviour for single solutes in defect-free Fe, as
discussed earlier. Indeed, experimental results\cite{Straalsund} find
Ni to be undersized and Cr oversized in an austenitic Fe-Cr-Ni alloy,
consistent with the pure elements. This may account for the earlier
attempts to suggest a positive binding of Ni to interstitial defects
which our results show is an unwarranted conclusion.

\begin{figure}
\includegraphics[width=\columnwidth]{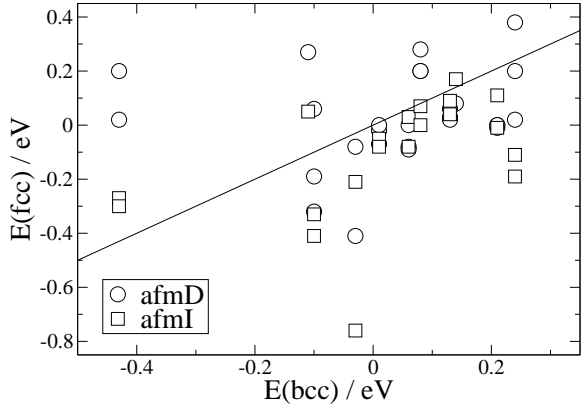}
\caption{\label{fccbccNiCrFig} Comparison of solute properties in
  austenitic and ferritic Fe. The dataset consists of the
  substitutional formation energy, solute-vacancy binding energies and
  solute-interstitial binding energies for the most stable
  self-interstitial, in eV. The black line is included to indicate an
  exact agreement.}
\end{figure}

As a final means to draw conclusions from the data we have plotted our
results for the solute substitution energy, solute-vacancy binding
energies and binding energies of solutes to the most stable
interstitial dumbbells in the mixed positions in austenitic Fe
against the corresponding values for bcc Fe
in \reffig{fccbccNiCrFig}. Ab initio data for bcc Fe was either as
published previously\cite{FeCrOlssonC,OlssonTMSol,KlaverFeCrBench} or
calculated here with similar settings.

It is immediately clear that there is no discernible correlation
between the two data sets, which is in complete contrast to the strong
similarity found for pure Fe. The (vertical) spread of fcc data also
illustrates just how strong the effects of symmetry breaking within
and between the two afm datasets can be, especially when compared to
the average binding energies. We note that data for the afmI state
appears to be lower, on average, than the afmD dataset. We suggest
that these result simply emphasise just how sensitive the interactions
of solutes and point defects are to their local magnetic environment
in the Fe-Cr-Ni system in addition to the usual volume-elastic
effects. The importance of developing a deeper understanding of these
effects cannot be overemphasised for the modelling of austenitic
alloys.

\section{Conclusions}
\label{conclusions}

We have carried out an extensive series of first-principles
calculations to determine the energetics of austenitic steels. We
first investigated a large set of possible reference states for
austenite at 0 K and found the fct afmI and afmD states to be the most
suitable, highlighting the problems with the unstable equilibrium
fm-HS states in the process. It would be incorrect to associate any
defect property of a paramagnetic material with a single
microstate. However, by sampling various reference configurations we
have been able to provide estimates of defect energies typically to
within a few tenths of 1 eV. The uncertainties associated with choice
of reference state in this work are in addition to the normal
discrepancy between ab initio and experimental quantities arising
from, on one hand, choice of pseudopotential, exchange-correlation
potential, neglect of zero-point energy, and on the other from finite
temperature and experimental errors. However, with considerably more
computational effort, our dataset for austenitic materials has only
moderately larger uncertainties than previous ones for ferritic
steels.

The main results and predictions of the paper can be summarised as
follows:

\begin{enumerate}[i/] 

\item The vacancy formation energy in austenite (at 0 K) is between
  1.8 and 1.95 eV. Divacancy binding is rather weak at around 0.1 eV,
  suggesting that the nucleation of voids will face a nucleation
  barrier at elevated temperatures. There is, however, a rapid
  increase in total binding energy with cluster size e.g.\ 2.5 eV for
  a six vacancy octahedral cluster. SFT exhibit similarly strong
  binding to the proto-voids (e.g.\ octahedral cluster) considered
  here and we therefore cannot distinguish between these two on
  energetic grounds.

\item The most stable self-interstitial is the $\langle 100\rangle$
  dumbbell, consistent with other fcc metals, with a formation energy
  of between 3.2 and 3.6 eV. These dumbbells aggregate strongly to
  form small defect clusters in (100) planes and although not
  considered here, pair binding also suggests that clusters in (111)
  planes would also exhibit high stability.

\item Pair bond models for the total binding of defect pairs and
  clusters agree well with the data and are able to capture and
  highlight general trends.

\item Our results show that Ni and Cr do not strongly attract each
  other: Binding energies of at most 0.1 eV were found for Ni-Ni
  pairs. The binding energies of solute clusters were consistent with
  pair interactions and fits to a pair bond model showed good
  agreement.

\item We find that Ni binds to vacancies but by no more than 0.1 eV
  and can therefore act as a weak trap for vacancy migration, which
  may be of importance for void formation and RIS in concentrated
  alloys. The Cr-vacancy interaction is weakly repulsive, suggesting
  that its effect on microstructure evolution and creep is negligible
  but concentration dependent effects cannot be ruled out in this
  study. Tracer diffusion coefficient calculations found that Ni
  diffuses significantly more slowly than Cr and Fe. Our calculations
  were also consistent with Cr diffusing in the opposite sense to the
  vacancy flux and indicate that Ni may diffuse with the vacancy flux.
  Both of these results are consistent with the standard mechanism
  used to explain the effects of RIS in austenitic alloys by vacancy
  mediated diffusion.

\item Cr was found to bind to mixed interstitial defects whereas Ni is
  generally repelled. The preferential association of a particular
  solute with the interstitial flux under irradiation is considered an
  important factor in RIS and our findings for Cr are worthy of
  consideration in the modelling of such effects. Substitutional Ni
  and Cr bind to $\langle 100\rangle$ self-interstitial dumbbells at
  nearest neighbour sites and therefore act as weak traps for their
  migration. 

\item Our results in austenitic Fe were compared with equivalent
  results in ferritic Fe. A strong similarity was found between these
  two datasets for point defect formation and binding energies in pure
  Fe. In contrast, no correlation was observed for Ni and Cr solute
  interactions with point defects.

\item We performed tests for the presence and influence of
  non-collinear magnetism in a subset of configurations. The tests
  either proved negative or resulted in marginal changes to the energy
  of the system. It should be borne in mind that such tests were
  useful to perform but are by no means exhaustive.

\end{enumerate}

In very broad terms, we have found that austenitic Fe behaves
similarly to other fcc metals, with $\langle 100\rangle$ interstitials
clustering to form proto-dislocation loops and vacancies clustering to
form sessile SFT and voids. This normality is in contrast to the often
anomalous behaviour of bcc Fe.  The interactions of Ni and Cr with
point defects are consistent with those of modestly oversized and
undersized defects, respectively, despite the experimentally observed
size factors and pure element data showing the opposite result.
Migration of Cr through the lattice by both vacancy and interstitial
mechanisms is enhanced relative to Fe self-diffusion.  By contrast Ni
migration is slow.  Pure fcc Fe is stable only at high temperature,
but the similarity to other fcc metals strongly suggests that FeCrNi
steels stabilised at lower temperature by alloying will also behave
normally.

Of course, to make these predictions more quantitative it will be
necessary to run further calculation using a method suited to larger
systems and longer timescales, such as molecular dynamics or kinetic
Monte Carlo. One of the main benefits of this study is to provide a
large database of configurations against which such models can be
parametrised.

\acknowledgements

This work was sponsored by the EU-FP7 PERFORM-60 project and by EPSRC
through the UKCP collaboration.

\appendix

\section{Tracer diffusion coefficients}
\label{appA}

The diffusion coefficient, $D$, describing atomic diffusion in a
crystalline material (in 3 dimensions) is given by\cite{LeClaire}
\be 
   D = \frac{1}{6} r^2 \Gamma f,
\ee
where $\Gamma$ is the number of atomic jumps per unit time, $r$ is the
length of each jump and $f$ is the correlation factor, which encodes
the fact that successive jumps are, in general, correlated. In a solid
solution the diffusion of the solvent element, A, and any
substitutional (i.e. on lattice) solute element, B, is mediated by
either vacancy or interstitial defects. In these cases it is clear
that correlations will arise because the defect mediating a jump will
still be present next to the migrating atom after the jump, which, for
example, makes the reverse jump more likely than others.

The tracer diffusion coefficient, $D^*_\mathrm{B}$, for a solute, B,
is the diffusion coefficient for B in solvent, A, in the limit where
the concentration of B atoms tends to zero. The jump frequency,
$\Gamma$, for vacancy-mediated tracer diffusion of B is given by
\be
   \Gamma = w_2 p_\mathrm{V},
\ee
where $w_2$ is the vacancy jump frequency for exchange with a B atom
at 1nn and $p_\mathrm{V}$ is the probability that a vacancy is
associated with a B atom at 1nn and is given by
\be
   p_\mathrm{V} = c_\mathrm{V} z \exp(\beta G_\mathrm{b,1nn}^\mathrm{B-V}),
\ee
where $c_\mathrm{V}$ is the vacancy concentration, z is the
coordination number of the lattice, $G_\mathrm{b,1nn}^\mathrm{B-V}$ is
the Gibbs free energy of binding for a B atom and vacancy at 1nn
(defined in an equivalent manner to the binding energy in
\eqn{EbindEq}) and $\beta = 1/k_\mathrm{B} T$ where $k_\mathrm{B}$ is
Boltzmann's constant and $T$ is the temperature. The tracer diffusion
coefficient is then given by
\be\label{Dstar}
   D^*_\mathrm{B} = \frac{1}{6} r^2 \omega_2 c_\mathrm{V} z \exp(\beta G_\mathrm{b,1nn}^\mathrm{B-V}) f_\mathrm{B}.
\ee
In the 5-frequency model of Lidiard and
LeClaire\cite{Lidiard,LeClaire} the correlation factor,
$f=f_\mathrm{B}$, is given in an fcc lattice by Manning\cite{Manning} as
\be\label{fBDefEq}
   f_\mathrm{B} = \frac{2w_1 + 7w_3F(w_4/w_0)}{2w_1 + 2w_2 + 7w_3F(w_4/w_0)},
\ee
where the $w_i$ are the vacancy jump frequencies for solvent-vacancy
exchanges: $w_1$ is where the vacancy is 1nn to B and remains so
after the jump; $w_3$ is where the vacancy is 1nn to B and does not
remain so after the jump i.e. a dissociative jump; $w_4$ is where the
vacancy is not at 1nn to B but is so after the jump i.e. an
associative jump and the opposite of $w_3$ and $w_0$ is for a jump in
the pure solvent i.e. self-diffusion. The factor, $F$, is the fraction
of dissociative jumps that do not return to a site 1nn to
B and we have used the expression from Koiwa and Ishioka\cite{Koiwa}
here
\be\label{KoiwaEq}
   7F(x) = 7 - \frac{a_1x+a_2x^2+a_3x^3+a_4x^4}{b_0+b_1x+b_2x^2+b_3x^3+b_4x^4},
\ee
where $x=w_4/w_0$ and the coefficients, $a_i$ and $b_i$ are given in \reftab{FcoeffsTab}.

\begin{table}[htbp]
\begin{ruledtabular}
\begin{tabular}{cccccc}
$i$ & 0 & 1 & 2 & 3 & 4 \\
\hline
$a_i$ & --- & 1338.0577 & 924.3303 & 180.3121 & 10 \\
$b_i$ & 435.2839 & 595.9725 & 253.3000 & 40.1478 & 2 \\
\end{tabular}
\end{ruledtabular}
\caption{\label{FcoeffsTab}Coefficients for \eqn{KoiwaEq} from the work of Koiwa and Ishioka\cite{Koiwa}.}
\end{table}

An identical expression for $D^*_\mathrm{B}$ has been derived by
Tucker {\it et al.}\cite{Tucker} from results for the
phenomenological coefficients by Allnatt\cite{Allnatt}.

The self-diffusion coefficient of solute A can be determined from
\eqn{Dstar} by considering B as a same-mass isotope of A. In this case
all jump frequencies equal $w_0$ and $G_\mathrm{b,1nn}^\mathrm{B-V} = 0$ giving
\be
   D^*_\mathrm{A} = \frac{1}{6}r^2w_0c_\mathrm{V}zf_0,
\ee
where the correlation factor, $f_0=0.7815$ for fcc.

We use an Arrhenius-type expression for the vacancy jump frequencies,
\be\label{freqEq}
   w_i = C_{\mathrm{m},i}\exp(-\beta H_{\mathrm{m},i}),
\ee
where $H_{\mathrm{m},i}$ is the enthalpy difference between the
transition state and initial (on-lattice) state at $T=0$ K for a $w_i$
jump i.e. the migration barrier height and $C_{\mathrm{m},i}$ is a
weakly temperature-dependent pre-factor. We include the zero-point
phonon contributions to the enthalpy at 0 K in the pre-factor. A
similar expression can be found for the binding energy factor in
\eqn{Dstar} by writing $G_\mathrm{b,1nn}^\mathrm{B-V}$ as a $T=0$ K
contribution, $H_\mathrm{b,1nn}^\mathrm{B-V}$, (which excludes the
zero-point energy) plus a term correcting for finite $T$, which gives
\be\label{bindEq}
   \exp(\beta G_\mathrm{b,1nn}^\mathrm{B-V}) = C_\mathrm{b} \exp(\beta H_\mathrm{b,1nn}^\mathrm{B-V}),
\ee
where the pre-factor, $C_\mathrm{b}$, contains the term correcting for
finite $T$ and is itself weakly temperature dependent.

Using \eqns{freqEq}{bindEq} we can write an expression for the ratio
of $D^*_\mathrm{B}$ to $D^*_\mathrm{A}$ as
\be\label{DstarRatio}
   R^\mathrm{B}_\mathrm{A} \equiv \frac{D^*_\mathrm{B}}{D^*_\mathrm{A}} = \frac{f_\mathrm{B}}{f_0} C_\mathrm{m} C_\mathrm{b} \exp(\beta H_{\mathrm{b},2}^\mathrm{B-TS}),
\ee
where
\be\label{CmigEq}
   C_\mathrm{m} = C_{\mathrm{m},2} / C_{\mathrm{m},0} 
\ee
and $H_{\mathrm{b},i}^\mathrm{B-TS}$ is the binding enthalpy for a B
solute to the transition state (TS) for a $w_i$ jump i.e. the lowering
of enthalpy resulting from exchanging a B atom infinitely far from a
migrating A atom with an A atom at the relevant site for a $w_i$
jump. It can be shown to be given by
\be\label{HeffEq}
   H_{\mathrm{b},i}^\mathrm{B-TS} = H_{\mathrm{m},0} - H_{\mathrm{m},i} + H_{\mathrm{b},i}^\mathrm{B-V},
\ee
where $H_{\mathrm{b},i}^\mathrm{B-V}$ is the binding enthalpy for the
initial (on-lattice) configuration for a $w_i$ jump. \eqn{HeffEq}
actually holds for any vacancy jump in the presence of a single B
solute atom and yields the correct limit that $H_{\mathrm{m},i}$ tends
to $H_{\mathrm{m},0}$ when the B atom is sufficiently far away that
the two binding energies are zero.

The relative diffusivities of B and A can be investigated using
\eqn{DstarRatio} as long as suitable expressions for all the factors
can be found.

The expression for the correlation factor, $f_\mathrm{B}$, can be
simplified using some reasonable approximations. First, note that we can
re-express \eqn{HeffEq} to give
\be
    H_{\mathrm{m},i} = H_{\mathrm{m},0} + H_{\mathrm{b},i}^\mathrm{B-V} - H_{\mathrm{b},i}^\mathrm{B-TS}.
\ee
For $w_3$ and $w_4$ jumps we make the assumption that
$H_{\mathrm{b},3}^\mathrm{B-TS} \equiv H_{\mathrm{b},4}^\mathrm{B-TS}
= 0$ i.e. that the presence of the B solute does not influence the
transition state enthalpy at this separation. This approximation is
consistent with the results presented for Cr and Fe solutes in fcc Ni
by Tucker {\it et al.}\cite{Tucker}, where migration enthalpies were
calculated directly, and we believe will also be for Cr and Ni solutes
in fcc Fe.

In addition, we make the approximation that $C_{\mathrm{m},i} =
C_{\mathrm{m},0}$ for all vacancy-solvent exchanges i.e. that this
factor only depends on the element being exchanged\cite{Tucker}. The
only other assumption we make is that $H_{\mathrm{b},4}^\mathrm{B-V} =
0$ i.e. that the binding between a B atom and vacancy is zero at 2nn,
3nn and 4nn separation, which is consistent with the data given in
\reftab{soluteVacTab}. Including these approximations in \eqn{fBDefEq}
gives
\be\label{fBapproxEq}
   f_\mathrm{B} = \frac{2\exp(\beta H_{\mathrm{b},1}^\mathrm{B-TS}) + 7F(1)}{2\exp(\beta  H_{\mathrm{b},1}^\mathrm{B-TS}) + 2 C_\mathrm{m}\exp(\beta  H_{\mathrm{b},2}^\mathrm{B-TS}) + 7F(1)},
\ee
which we use along with \eqn{DstarRatio} to investigate the relative
diffusivities of Ni and Cr in fcc Fe.

\section{Vacancy Wind}
\label{appB}

It is also useful when investigating defect-mediated diffusion to be
able to determine the direction of the solute flux relative to that of
the corresponding defect flux. For vacancy-mediated diffusion the
vacancy wind\cite{ManningWind}, $G$, allows this relationship to be
investigated. When $G > -1$ the flux of B solute atoms is opposite to
the vacancy flux. By contrast, when $G < -1$ they are in the same
direction and the vacancy flux tends to drag any solute along with it.

\begin{figure}
\includegraphics[width=\columnwidth]{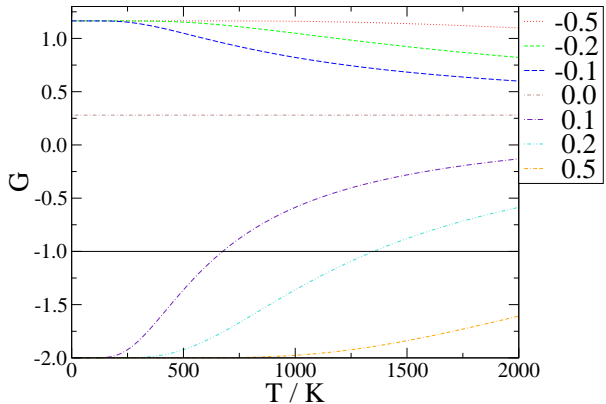}
\caption{\label{vacancyWindFig}The vacancy wind, $G$, versus
  temperature, T, for a number of distinct values for
  $H_{\mathrm{b},1}^\mathrm{B-TS}$, which are used to label the
  corresponding curves. The line $G=-1$ is shown to distinguish the
  vacancy drag regime from the regime where solute flux is opposite to
  vacancy flux.}
\end{figure}

In the 5-frequency model for fcc the vacancy wind is given by
\be
   G = \frac{6w_3 - 4w_1 + 14w_3(1-F(w_4/w_0))(w_0-w_4)/w_4}{2w_1+7w_3F(w_4/w_0)},
\ee
which with the approximations used here becomes
\be\label{vacancyWindEq}
   G = \frac{6-4\exp(\beta H_{\mathrm{b},1}^\mathrm{B-TS})}{2\exp(\beta H_{\mathrm{b},1}^\mathrm{B-TS}) + 7F(1)}.
\ee
The temperature dependence of $G$ is controlled by the single
parameter, $H_{\mathrm{b},1}^\mathrm{B-TS}$, which essentially
determines whether $w_1$ vacancy jumps are more probable (for positive
values) or less probable (for negative values) than dissociative $w_3$
jumps. When $w_1$ jumps are significantly more probable diffusion is
dominated by the Johnson mechanism\cite{JohnsonMechanism} where
vacancy-solute complexes diffuse cooperatively. We plot $G$ versus
temperature in \reffig{vacancyWindFig} for a number of values of
$H_{\mathrm{b},1}^\mathrm{B-TS}$.

It is clear that if $H_{\mathrm{b},1}^\mathrm{B-TS} \le 0$ then $G>-1$
and the solute flux is opposite the vacancy flux for all
temperatures. However, if $H_{\mathrm{b},1}^\mathrm{B-TS} > 0$, then
there exists a critical temperature, below which the vacancy drag
mechanism prevails and diffusion is primarily by the Johnson
mechanism.

\end{document}